\documentclass[12pt]{article}
\usepackage{amssymb}
\usepackage{color,graphicx}
\usepackage{amsmath}
\usepackage[v1compatible]{axodraw2}

\newcommand{\ep}{\varepsilon}
\newcommand{\bea}{\begin{eqnarray}}
\newcommand{\eea}{\end{eqnarray}}
\newcommand{\be}{\begin{equation}}
\newcommand{\ee}{\end{equation}}

\begin{document}
%\selectlanguage{english}

\begin{center}
  {\Large {\bf
      Differential equations and
      %in the calculation of
      Feynman integrals
%Differential equations in action
}}
\\ \vspace*{5mm} A.~V.~Kotikov
%$^{a}$
\end{center}

\begin{center}
%${}^{a}$
Bogoliubov Laboratory of Theoretical Physics \\
Joint Institute for Nuclear Research\\
141980 Dubna, Russia
\end{center}

\begin{abstract}
%\noindent
%  We report
%  %  review
%  some results of calculations of
%  %massless and
%  massive Feynman integrals
%  particularly focusing on difference equations for coefficients of for their series expansions

  The role of differential equations in the process of calculating
  %massive
  Feynman integrals is reviewed.
%  shown.
  An example of a diagram is given
  %, in the calculation of
  for which the method of differential equations was introduced,
the properties of the inverse-mass-expansion coefficients are shown, and modern methods based on differential
equations are considered.
  
\end{abstract}

% \pacs{12.38.Bx}
%\maketitle
%{\em PACS:} 12.38.Bx

\section{Introduction}
The calculation of the Feynman integrals (FIs) provides basic information both for the matrix elements of the experimentally
studied processes and for the characteristics of the physical models themselves, i.e. their renormalization, critical behavior, etc.
When studying renormalization and the critical behavior, it is usually sufficient to restrict oneself to the limit of massless particles
at which the corresponding two-point FIs are fairly simple. However, starting at 2 or 3 loop level, there is a need to use modern methods
such as integration by parts (IBP)  \cite{Chetyrkin:1981qh} and the Gegenbauer's polynomial method \cite{Chetyrkin:1980pr}.
\footnote{See also Ref.
  %the recent paper
  \cite{Kotikov:2013eha} and the reviews   \cite{Grozin:2005yg} and \cite{Teber:2016unz}).
  %\footnote{
  Note that multipoint massless FIs are as complex as massive 2-point FIs. For the 
%  (see, for example, Ref. \cite{Davydychev:1995mq} about the
  relationship between 2-point massive FIs and 3-point massless FIs, cf.  \cite{Davydychev:1995mq} .} 

Calculating FIs having
%propagators of
massive propagators
is a much more complex problem. Simple results, in the form of a product of $\Gamma$-functions exist 
for simple tadpoles only, see Eq. (\ref{Tp}) below. A massive one-loop loop is already given by a one-fold
%single
integral, see Eq. (\ref{loopM}) below.

It turned out, however, that massive FIs satisfy IBP procedures  \cite{Chetyrkin:1981qh}, which lead to relations between the FIs
equivalent to the original ones,
but with different powers of the propagators, including powers equal to zero. Diagrams containing propagators with
degrees equal to zero are equivalent to simpler diagrams obtained by canceling these propagators and reducing the points they join
to one point.

Such relations can be understood in two ways.
First, considering them algebraically, one can understand them as connections between diagrams that are not independent and can be
reduced to a certain set of independent diagrams, which are called master integrals (or masters) \cite{Broadhurst:1987ei}.

Second, propagators with powers greater than one can be considered as derivatives, with respect to the corresponding mass or external
momentum, from the propagator with a degree of one.
Thus, the relations between the master integrals can be considered as differential equations (DEs) for these masters. An example is given in
Section 2, containing inhomogeneous terms, including only simpler diagrams, which are obtained from the original diagrams by reducing
some propagator.
For these simpler diagrams one can obtain
%(by applying the IBP procedure to them)
similar DEs by applying the IBP procedure,
%differential equations
see the Appendix. They
contain inhomogeneous terms, including only even simpler diagrams, which are obtained from simple diagrams  by propagator reduction.
%some propagator).
By repeating the original procedure several times, it is usually possible to obtain DEs
%differential equations
containing inhomogeneous terms, including only tadpoles, which in turn are easily computable exactly.
Note, however, that starting from the 2-loop level, obtaining results for massive tadpoles requires the use of modern methods of FI calculation, cf.
%(see
%the recent papers
\cite{Kniehl:2017ikj} and references and discussions therein.
%\footnote{
Sometimes it is convenient to stop the considered procedure on one-loop massive FI and to perform  the integration after introducing
%calculation of which can be performed,for example, by the method of
Feynman parameters, cf. \cite{Ryder}.
More complicated diagrams can be obtained from these tadpoles by solving successively obtained DEs
%differential equations
with certain boundary
conditions. For dimensionally regularized massive FIs a good boundary condition is obtained in the limit of
%infinitely
large masses, $m \to \infty$,
at which these diagrams usually vanish.

The paper is organized as follows.
In Section 2 we will consider a two-loop FI, the calculation of which leds to the use of differential equations.
The calculation of massive diagrams is given in Section 3. Here rules are given for their efficient
calculation, examples of two- and three-point diagrams are considered. The recurrence relations for the coefficients
of decomposition in the inverse mass are considered.
In Section 4 a short review of modern computing technology is given.
The appendix  contains the derivation of the DEs
%differential equations
for massive diagrams from the inhomogeneous term of the DE
%differential equation
for the diagram considered in Section 2.

%%%%%%%%%%%%%%%%% NEW %%%%%%%%%%%
\section{History}

As mentioned in the introduction, integral representations for one-loop FIs (obtained, for example, using the Feynman
parameter method \cite{Ryder}) are hypergeometric functions\footnote{Investigations of hypergeometric functions related to the calculation of FI are recently presented \cite{Kalmykov:2020cqz}
  as contribution to this volume.}
and, thus, can be represented as solutions of some DEs.
%differential equations.
The importance of DEs
%differential equations
for FIs was recognized long ago, see, for example, \cite{Regge,Golubeva}. However, in my
opinion, the practical application
awaited the emergence of the IBP procedure  \cite{Chetyrkin:1981qh} for FIs and is based on the use of IBP relations, see Eqs.
(\ref{TreIBPM}) and (\ref{IBPpro}) below.

IBP-based DEs
%differential equations
appeared in the nineties in several works, studying FIs: for massive two-point functions in
\cite{Kotikov:1990kg,Remiddi:1997ny}, for massive three-point functions in \cite{Kotikov:1991hm}, and for four-point  in
\cite{Gehrmann:1999as}. Also $n$-point functions were considered in \cite{Kotikov:1991pm,Bern:1994zx}.
A short overview was given in Ref.
%made at that time can be found in the issue
\cite{Kotikov:1991zd} dedicated to the 70th anniversary of Academician
O.S. Parasyuk, the co-author of the BPHZ renormalization procedure  \cite{Bogoliubov:1957gp}, cf. e.g.
%(see, for example, the book
\cite{Bogolyubov:1980nc}.
%\footnote{
The results for massless diagrams are sometimes obtained more easily in $x$-space, cf. \cite{Kazakov:1984km,Kazakov:1986mu,Gracey:2013sz}.
It is convenient to ccompute
%do this in
%%  Contrary to original papers \cite{Kazakov:1986mu,Kotikov:1987mw} all calculations are carried out
%%in $p$-space. We note that original results have been done in
%$x$-space for
the so-called dual diagrams in $x$-space, cf.
%(see, for example,
\cite{Kazakov:1986mu,Kazakov:1987jk}. A dual diagram is obtained
from the initial one by replacement of all momenta $p$ by $x$ with the rules of correspondence between the
graph and the integral, as in a $x$-space. Massive two-pint and three-point diagrams were studied in dual $x$-space 
in Refs. \cite{Kotikov:1990zs} and \cite{Kotikov:1990zk}, respectively.
%The transition to the dual diagram is indicated by $\stackrel{d}{=}$.}DEM2

In Refs. \cite{Kotikov:1990kg,Kotikov:1990zs}
%, the starting point was to
we studied a preprint of the excellent yet
unpublished work  \cite{Broadhurst:1987ei} on the calculation of two-loop massive FIs. Despite the excellent results,
the paper itself
%(i.e. the conclusions of these results)
turned out to be quite difficult to understand.

I therefore decided to reproduce these results using the IBP relations, which proved to be very successful for calculating the correction to
the longitudinal structure function of the deep-inelastic scattering (DIS) \cite{Kazakov:1987jk,KK92}. Indeed, the method developed
 \cite{Kazakov:1986mu,Kazakov:1987jk} for calculating massless FIs containing the (traceless) product of impulses in
the numerators of propagators was based on the application of IBP to such diagrams.
This method, extended to 3-, 4- and 5-loop diagrams and built into computer algebra programs, is the basis of the modern calculations,
starting with the excellent work in which NNLO corrections for anomalous dimensions of Wilson operators were obtained,
%(a recent overview of the results can be found in
see e.g.
%the recent paper
\cite{Herzog:2018kwj}, and references and discussions therein. A similar method has also been developed 
\cite{Blumlein:2009rg} to
calculate massive corrections in the DIS process, cf.
%(see the recent papers
\cite{Behring:2021asx,Ablinger:2020snj}
and the review \cite{Blumlein:2018cms} and details given
%references and discussions
therein.

The first example which was studied in Refs.  \cite{Kotikov:1990kg,Kotikov:1990zs} was the diagram
\vskip 0.5cm
\be
I_1(q^2,m^2) \, =
 \hspace{3mm}
%\frac{2^{n-m} \Gamma(n-m+\alpha)}{(n-m)!\Gamma(\alpha)} \, \hspace{3mm}
\raisebox{1mm}{{
    %\begin{picture}(90,30)(0,4)
    \begin{axopicture}(90,10)(0,4)
%  \SetWidth{2.0}
  \SetWidth{0.5}
%\CArc(5,5)(80,20,160)
%\CArc(45,5)(40,0,180)
%\Arc(45,-7)(40,20,160)
\Arc(45,-7)(40,20,90)
%\Arc[arrow](45,-7)(40,90,160)
\Arc(45,-7)(40,90,160)
 \SetWidth{1.5}
\Line(45,-25)(45,35)
 \SetWidth{0.5}
%\Line[arrow](40,5)(85,5)
\Arc(45,17)(40,200,270)
%\Arc[arrow](45,17)(40,270,340)
\Arc(45,17)(40,270,340)
%\Arc(45,17)(40,200,340)
\Vertex(45,-23){2}
\Vertex(45,33){2}
\Vertex(5,5){2}
\Vertex(85,5){2}
%\SetWidth{1.0}
%\Vertex(5,5){2}
\Line(5,5)(-5,5)
\Line(85,5)(95,5)
%Line(45,-25)(45,-35)
%Line(85,5)(95,10)
%\Vertex(40,5){2}
%\Vertex(40,15){2}
%\Vertex(40,-15){2}
%\Text(25,33)[b]{$m$}
%\Text(63,-10)[b]{$\scriptstyle n$}
%\Text(63,-10)[b]{$n$}
%\Text(33,3)[t]{$\scriptstyle \frac{M^2}{s(1-s)}$}
%\Text(33,1)[t]{$\scriptstyle M^2/[s(1-s)]$}
%\Text(75,-22)[t]{$\alpha$}
%\Text(75,-22)[t]{$\scriptstyle \alpha$}
\Text(-3,-5)[b]{$\to$}
\Text(-3,-12)[b]{$q$}
%\Text(93,-5)[b]{$\to$}
%\Text(93,-12)[b]{$q$}
%\Text(93,20)[b]{$\to$}
%\Text(93,12)[b]{$p$}
%\Text(52,-30)[b]{$\to$}
%\Text(52,-38)[b]{$p$}
%\Text(-3,-5)[b]{$p$}
\end{axopicture}
}} \, ,
\label{I1}
\ee
\vskip 1cm
%,
\noindent
having the vertical massive propagator, see Eq. (\ref{DefM}) for definitions. The diagram 
has left-right and top-bottom symmetries.

Applying IBP relations (\ref{TreIBPM}) to the left triangle of the diagram $I_1(q^2,m^2)$
in succession with vertical and lateral distinguished lines, we get
\bea
&&(d-4) I_1(q^2,m^2) = 2 \Biggl[\hspace{3mm}
%\small{
  \raisebox{1mm}{{
    %\begin{picture}(90,30)(0,4) 
    \begin{axopicture}(90,10)(0,4)
%  \SetWidth{2.0}
  \SetWidth{0.5}
\CArc(25,5)(20,0,180)
\CArc(25,5)(20,180,360)
%  \Arc(20,-7)(20,20,160)
%  \Arc(20,17)(20,200,340)
%\Line[arrow](5,5)(40,5)
%\Line[arrow](40,5)(85,5)
 \Arc(65,5)(20,0,180)
  \Arc(65,5)(20,180,360)
  \Vertex(5,5){2}
  \Vertex(45,5){2}
\Vertex(85,5){2}
%\SetWidth{1.0}
%\Vertex(5,5){2}
\Line(5,5)(-5,5)
\Line(85,5)(95,5)
%\Vertex(40,5){2}
%\Vertex(40,15){2}
%\Vertex(40,-15){2}
%\Text(65,-10)[b]{$n$}
%\Text(65,10)[b]{$m$}
%\Text(33,3)[t]{$\scriptstyle \frac{M^2}{s(1-s)}$}
%\Text(33,1)[t]{$\scriptstyle M^2/[s(1-s)]$}
%\Text(45,27)[t]{$\alpha_1$}
%\Text(65,-20)[t]{$\alpha$}
\Text(25,-20)[t]{$2$}
\Text(-3,-5)[b]{$\to$}
\Text(-3,-12)[b]{$q$}
%\Text(-3,-5)[b]{$p$}
\end{axopicture}
  }}
%  }
  \hspace{3mm}
  -
    \raisebox{1mm}{{
    %\begin{picture}(90,30)(0,4)
    \begin{axopicture}(90,10)(0,4)
%  \SetWidth{2.0}
  \SetWidth{0.5}
%\CArc(5,5)(80,20,160)
%\CArc(45,5)(40,0,180)
\Arc(45,-7)(40,20,90)
\Arc(45,-7)(40,90,160)
%\Line[arrow](5,5)(40,5)
%\Line[arrow](40,5)(85,5)
%\Arc(5,-35)(40,20,85)
%\Arc[arrow](85,45)(40,197,270)
\SetWidth{1.5}
\Arc(85,45)(40,197,270)
\SetWidth{0.5}
\Arc(45,17)(40,200,270)
\Arc(45,17)(40,270,340)
\Vertex(5,5){2}
\Vertex(85,5){2}
\Vertex(48,33){2}
%\SetWidth{1.0}
%\Vertex(5,5){2}
\Line(5,5)(-5,5)
\Line(85,5)(95,5)
%\Vertex(40,5){2}
%\Vertex(40,15){2}
%\Vertex(40,-15){2}
%\Text(67,13)[b]{$\scriptstyle k$}
%\Text(65,10)[b]{$m$}
%\Text(33,3)[t]{$\scriptstyle \frac{M^2}{s(1-s)}$}
%\Text(33,1)[t]{$\scriptstyle M^2/[s(1-s)]$}
%\Text(65,-25)[t]{$\alpha$}
%\Text(33,7)[t]{$2$}
\Text(25,-20)[t]{$2$}
\Text(-3,-5)[b]{$\to$}
\Text(-3,-12)[b]{$q$}
%\Text(-3,-5)[b]{$p$}
\end{axopicture}
  }} \hspace{0.5cm} \nonumber \\
  %\Biggr)
&&\nonumber \\ && \nonumber \\ && \nonumber \\
    && \hspace{2.5cm} - m^2 \,
   \hspace{3mm}
%\frac{2^{n-m} \Gamma(n-m+\alpha)}{(n-m)!\Gamma(\alpha)} \, \hspace{3mm}
\raisebox{1mm}{{
    %\begin{picture}(90,30)(0,4)
    \begin{axopicture}(90,10)(0,4)
%  \SetWidth{2.0}
  \SetWidth{0.5}
%\CArc(5,5)(80,20,160)
%\CArc(45,5)(40,0,180)
%\Arc(45,-7)(40,20,160)
\Arc(45,-7)(40,20,90)
%\Arc[arrow](45,-7)(40,90,160)
\Arc(45,-7)(40,90,160)
 \SetWidth{1.5}
\Line(45,-25)(45,35)
 \SetWidth{0.5}
%\Line[arrow](40,5)(85,5)
\Arc(45,17)(40,200,270)
%\Arc[arrow](45,17)(40,270,340)
\Arc(45,17)(40,270,340)
%\Arc(45,17)(40,200,340)
\Vertex(45,-23){2}
\Vertex(45,33){2}
\Vertex(5,5){2}
\Vertex(85,5){2}
%\SetWidth{1.0}
%\Vertex(5,5){2}
\Line(5,5)(-5,5)
\Line(85,5)(95,5)
%Line(45,-25)(45,-35)
%Line(85,5)(95,10)
%\Vertex(40,5){2}
%\Vertex(40,15){2}
%\Vertex(40,-15){2}
%\Text(25,33)[b]{$m$}
%\Text(63,-10)[b]{$\scriptstyle n$}
%\Text(63,-10)[b]{$n$}
%\Text(33,3)[t]{$\scriptstyle \frac{M^2}{s(1-s)}$}
%\Text(33,1)[t]{$\scriptstyle M^2/[s(1-s)]$}
%\Text(75,-22)[t]{$\alpha$}
%\Text(75,-22)[t]{$\scriptstyle \alpha$}
\Text(-3,-5)[b]{$\to$}
\Text(-3,-12)[b]{$q$}
\Text(25,-20)[t]{$2$}
%\Text(93,-5)[b]{$\to$}
%\Text(93,-12)[b]{$q$}
%\Text(93,20)[b]{$\to$}
%\Text(93,12)[b]{$p$}
%\Text(52,-30)[b]{$\to$}
%\Text(52,-38)[b]{$p$}
%\Text(-3,-5)[b]{$p$}
\end{axopicture}
}}
\hspace{3mm}
\Biggr]
- 2 m^2 \,
   \hspace{3mm}
%\frac{2^{n-m} \Gamma(n-m+\alpha)}{(n-m)!\Gamma(\alpha)} \, \hspace{3mm}
\raisebox{1mm}{{
    %\begin{picture}(90,30)(0,4)
    \begin{axopicture}(90,10)(0,4)
%  \SetWidth{2.0}
  \SetWidth{0.5}
%\CArc(5,5)(80,20,160)
%\CArc(45,5)(40,0,180)
%\Arc(45,-7)(40,20,160)
\Arc(45,-7)(40,20,90)
%\Arc[arrow](45,-7)(40,90,160)
\Arc(45,-7)(40,90,160)
 \SetWidth{1.5}
\Line(45,-25)(45,35)
 \SetWidth{0.5}
%\Line[arrow](40,5)(85,5)
\Arc(45,17)(40,200,270)
%\Arc[arrow](45,17)(40,270,340)
\Arc(45,17)(40,270,340)
%\Arc(45,17)(40,200,340)
\Vertex(45,-23){2}
\Vertex(45,33){2}
\Vertex(5,5){2}
\Vertex(85,5){2}
%\SetWidth{1.0}
%\Vertex(5,5){2}
\Line(5,5)(-5,5)
\Line(85,5)(95,5)
%Line(45,-25)(45,-35)
%Line(85,5)(95,10)
%\Vertex(40,5){2}
%\Vertex(40,15){2}
%\Vertex(40,-15){2}
%\Text(25,33)[b]{$m$}
%\Text(63,-10)[b]{$\scriptstyle n$}
%\Text(63,-10)[b]{$n$}
%\Text(33,3)[t]{$\scriptstyle \frac{M^2}{s(1-s)}$}
%\Text(33,1)[t]{$\scriptstyle M^2/[s(1-s)]$}
%\Text(75,-22)[t]{$\alpha$}
%\Text(75,-22)[t]{$\scriptstyle \alpha$}
\Text(-3,-5)[b]{$\to$}
\Text(-3,-12)[b]{$q$}
\Text(52,5)[t]{$2$}
%\Text(93,-5)[b]{$\to$}
%\Text(93,-12)[b]{$q$}
%\Text(93,20)[b]{$\to$}
%\Text(93,12)[b]{$p$}
%\Text(52,-30)[b]{$\to$}
%\Text(52,-38)[b]{$p$}
%\Text(-3,-5)[b]{$p$}
\end{axopicture}
}} \hspace{3mm} \, ,
\label{eq1} \\
&&\nonumber \\ && \nonumber \\ && \nonumber \\
&&(d-4) I_1(q^2,m^2) =
\hspace{3mm}
%\nonumber \\ &&
%\nonumber \\ &&
%\nonumber \\ &&
%\nonumber \\ &&
  \raisebox{1mm}{{
    %\begin{picture}(90,30)(0,4)
    \begin{axopicture}(90,10)(0,4)
%  \SetWidth{2.0}
  \SetWidth{0.5}
%\CArc(5,5)(80,20,160)
%\CArc(45,5)(40,0,180)
\Arc(45,-7)(40,20,160)
  %\Line[arrow](5,5)(40,5)
%\Line[arrow](40,5)(85,5)
\SetWidth{1.5}
\Arc(5,-35)(40,20,85)
\SetWidth{0.5}
\Arc(45,17)(40,200,270)
\SetWidth{0.5}
\Arc(45,17)(40,270,340)
\SetWidth{0.5}
\Vertex(5,5){2}
\Vertex(85,5){2}
\Vertex(43,-23){2}
%\SetWidth{1.0}
%\Vertex(5,5){2}
\Line(5,5)(-5,5)
\Line(85,5)(95,5)
%\Vertex(40,5){2}
%\Vertex(40,15){2}
%\Vertex(40,-15){2}
%\Text(65,-10)[b]{$n$}
%\Text(65,10)[b]{$m$}
%\Text(33,3)[t]{$\scriptstyle \frac{M^2}{s(1-s)}$}
%\Text(33,1)[t]{$\scriptstyle M^2/[s(1-s)]$}
%\Text(65,-25)[t]{$\alpha-1$}
%\Text(33,7)[t]{$2$}
\Text(25,-20)[t]{$2$}
\Text(-3,-5)[b]{$\to$}
\Text(-3,-12)[b]{$q$}
%\Text(-3,-5)[b]{$p$}
\end{axopicture}
  }}
% \nonumber \\
%  %\Biggr)
%&&\nonumber \\ && \nonumber \\ && \nonumber \\
  %
   \hspace{0.5cm}
  - q^2 \,
   \hspace{3mm}
%\frac{2^{n-m} \Gamma(n-m+\alpha)}{(n-m)!\Gamma(\alpha)} \, \hspace{3mm}
\raisebox{1mm}{{
    %\begin{picture}(90,30)(0,4)
    \begin{axopicture}(90,10)(0,4)
%  \SetWidth{2.0}
  \SetWidth{0.5}
%\CArc(5,5)(80,20,160)
%\CArc(45,5)(40,0,180)
%\Arc(45,-7)(40,20,160)
\Arc(45,-7)(40,20,90)
%\Arc[arrow](45,-7)(40,90,160)
\Arc(45,-7)(40,90,160)
 \SetWidth{1.5}
\Line(45,-25)(45,35)
 \SetWidth{0.5}
%\Line[arrow](40,5)(85,5)
\Arc(45,17)(40,200,270)
%\Arc[arrow](45,17)(40,270,340)
\Arc(45,17)(40,270,340)
%\Arc(45,17)(40,200,340)
\Vertex(45,-23){2}
\Vertex(45,33){2}
\Vertex(5,5){2}
\Vertex(85,5){2}
%\SetWidth{1.0}
%\Vertex(5,5){2}
\Line(5,5)(-5,5)
\Line(85,5)(95,5)
%Line(45,-25)(45,-35)
%Line(85,5)(95,10)
%\Vertex(40,5){2}
%\Vertex(40,15){2}
%\Vertex(40,-15){2}
%\Text(25,33)[b]{$m$}
%\Text(63,-10)[b]{$\scriptstyle n$}
%\Text(63,-10)[b]{$n$}
%\Text(33,3)[t]{$\scriptstyle \frac{M^2}{s(1-s)}$}
%\Text(33,1)[t]{$\scriptstyle M^2/[s(1-s)]$}
%\Text(75,-22)[t]{$\alpha$}
%\Text(75,-22)[t]{$\scriptstyle \alpha$}
\Text(-3,-5)[b]{$\to$}
\Text(-3,-12)[b]{$q$}
\Text(25,-20)[t]{$2$}
%\Text(93,-5)[b]{$\to$}
%\Text(93,-12)[b]{$q$}
%\Text(93,20)[b]{$\to$}
%\Text(93,12)[b]{$p$}
%\Text(52,-30)[b]{$\to$}
%\Text(52,-38)[b]{$p$}
%\Text(-3,-5)[b]{$p$}
\end{axopicture}
}}
\hspace{3mm}
\nonumber \\ && \nonumber \\ && \nonumber \\&& \nonumber \\
%&& \nonumber \\
&&\hspace{2.5cm} -  m^2 \,
   \hspace{3mm}
%\frac{2^{n-m} \Gamma(n-m+\alpha)}{(n-m)!\Gamma(\alpha)} \, \hspace{3mm}
\raisebox{1mm}{{
    %\begin{picture}(90,30)(0,4)
    \begin{axopicture}(90,10)(0,4)
%  \SetWidth{2.0}
  \SetWidth{0.5}
%\CArc(5,5)(80,20,160)
%\CArc(45,5)(40,0,180)
%\Arc(45,-7)(40,20,160)
\Arc(45,-7)(40,20,90)
%\Arc[arrow](45,-7)(40,90,160)
\Arc(45,-7)(40,90,160)
 \SetWidth{1.5}
\Line(45,-25)(45,35)
 \SetWidth{0.5}
%\Line[arrow](40,5)(85,5)
\Arc(45,17)(40,200,270)
%\Arc[arrow](45,17)(40,270,340)
\Arc(45,17)(40,270,340)
%\Arc(45,17)(40,200,340)
\Vertex(45,-23){2}
\Vertex(45,33){2}
\Vertex(5,5){2}
\Vertex(85,5){2}
%\SetWidth{1.0}
%\Vertex(5,5){2}
\Line(5,5)(-5,5)
\Line(85,5)(95,5)
%Line(45,-25)(45,-35)
%Line(85,5)(95,10)
%\Vertex(40,5){2}
%\Vertex(40,15){2}
%\Vertex(40,-15){2}
%\Text(25,33)[b]{$m$}
%\Text(63,-10)[b]{$\scriptstyle n$}
%\Text(63,-10)[b]{$n$}
%\Text(33,3)[t]{$\scriptstyle \frac{M^2}{s(1-s)}$}
%\Text(33,1)[t]{$\scriptstyle M^2/[s(1-s)]$}
%\Text(75,-22)[t]{$\alpha$}
%\Text(75,-22)[t]{$\scriptstyle \alpha$}
\Text(-3,-5)[b]{$\to$}
\Text(-3,-12)[b]{$q$}
\Text(52,5)[t]{$2$}
%\Text(93,-5)[b]{$\to$}
%\Text(93,-12)[b]{$q$}
%\Text(93,20)[b]{$\to$}
%\Text(93,12)[b]{$p$}
%\Text(52,-30)[b]{$\to$}
%\Text(52,-38)[b]{$p$}
%\Text(-3,-5)[b]{$p$}
\end{axopicture}
}} \hspace{3mm} \, .
\label{eq2}
\eea
\vskip 1cm

Taking the combination of these equations: Eq. (\ref{eq1}) - 2$(m^2/q^2) \times$ Eq.(\ref{eq2}), we have
\vskip 0.5cm
\be
(d-4) \, \left(1-\frac{2m^2}{q^2}\right) \, I_1(q^2,m^2) = 2 \, J_1(q^2,m^2) -
2m^2 \, \left(1-\frac{m^2}{q^2}\right) \, 
 \hspace{3mm}
%\frac{2^{n-m} \Gamma(n-m+\alpha)}{(n-m)!\Gamma(\alpha)} \, \hspace{3mm}
\raisebox{1mm}{{
    %\begin{picture}(90,30)(0,4)
    \begin{axopicture}(90,10)(0,4)
%  \SetWidth{2.0}
  \SetWidth{0.5}
%\CArc(5,5)(80,20,160)
%\CArc(45,5)(40,0,180)
%\Arc(45,-7)(40,20,160)
\Arc(45,-7)(40,20,90)
%\Arc[arrow](45,-7)(40,90,160)
\Arc(45,-7)(40,90,160)
 \SetWidth{1.5}
\Line(45,-25)(45,35)
 \SetWidth{0.5}
%\Line[arrow](40,5)(85,5)
\Arc(45,17)(40,200,270)
%\Arc[arrow](45,17)(40,270,340)
\Arc(45,17)(40,270,340)
%\Arc(45,17)(40,200,340)
\Vertex(45,-23){2}
\Vertex(45,33){2}
\Vertex(5,5){2}
\Vertex(85,5){2}
%\SetWidth{1.0}
%\Vertex(5,5){2}
\Line(5,5)(-5,5)
\Line(85,5)(95,5)
%Line(45,-25)(45,-35)
%Line(85,5)(95,10)
%\Vertex(40,5){2}
%\Vertex(40,15){2}
%\Vertex(40,-15){2}
%\Text(25,33)[b]{$m$}
%\Text(63,-10)[b]{$\scriptstyle n$}
%\Text(63,-10)[b]{$n$}
%\Text(33,3)[t]{$\scriptstyle \frac{M^2}{s(1-s)}$}
%\Text(33,1)[t]{$\scriptstyle M^2/[s(1-s)]$}
%\Text(75,-22)[t]{$\alpha$}
%\Text(75,-22)[t]{$\scriptstyle \alpha$}
\Text(-3,-5)[b]{$\to$}
\Text(-3,-12)[b]{$q$}
\Text(52,5)[t]{$2$}
%\Text(93,-5)[b]{$\to$}
%\Text(93,-12)[b]{$q$}
%\Text(93,20)[b]{$\to$}
%\Text(93,12)[b]{$p$}
%\Text(52,-30)[b]{$\to$}
%\Text(52,-38)[b]{$p$}
%\Text(-3,-5)[b]{$p$}
\end{axopicture}
}} \hspace{3mm} \, ,
\label{eq3}
\ee
\vskip 1cm
\noindent
where
\be
J_1(q^2,m^2) =  
\hspace{3mm}
  \raisebox{1mm}{{
    %\begin{picture}(90,30)(0,4) 
    \begin{axopicture}(90,10)(0,4)
%  \SetWidth{2.0}
  \SetWidth{0.5}
\CArc(25,5)(20,0,180)
\CArc(25,5)(20,180,360)
%  \Arc(20,-7)(20,20,160)
%  \Arc(20,17)(20,200,340)
%\Line[arrow](5,5)(40,5)
%\Line[arrow](40,5)(85,5)
 \Arc(65,5)(20,0,180)
  \Arc(65,5)(20,180,360)
  \Vertex(5,5){2}
  \Vertex(45,5){2}
\Vertex(85,5){2}
%\SetWidth{1.0}
%\Vertex(5,5){2}
\Line(5,5)(-5,5)
\Line(85,5)(95,5)
%\Vertex(40,5){2}
%\Vertex(40,15){2}
%\Vertex(40,-15){2}
%\Text(65,-10)[b]{$n$}
%\Text(65,10)[b]{$m$}
%\Text(33,3)[t]{$\scriptstyle \frac{M^2}{s(1-s)}$}
%\Text(33,1)[t]{$\scriptstyle M^2/[s(1-s)]$}
%\Text(45,27)[t]{$\alpha_1$}
%\Text(65,-20)[t]{$\alpha$}
\Text(25,-20)[t]{$2$}
\Text(-3,-5)[b]{$\to$}
\Text(-3,-12)[b]{$q$}
%\Text(-3,-5)[b]{$p$}
\end{axopicture}
}}
  \hspace{3mm}
  -
\hspace{3mm}
  \raisebox{1mm}{{
    %\begin{picture}(90,30)(0,4)
    \begin{axopicture}(90,10)(0,4)
%  \SetWidth{2.0}
  \SetWidth{0.5}
%\CArc(5,5)(80,20,160)
%\CArc(45,5)(40,0,180)
\Arc(45,-7)(40,20,90)
\Arc(45,-7)(40,90,160)
%\Line[arrow](5,5)(40,5)
%\Line[arrow](40,5)(85,5)
%\Arc(5,-35)(40,20,85)
%\Arc[arrow](85,45)(40,197,270)
\SetWidth{1.5}
\Arc(85,45)(40,197,270)
\SetWidth{0.5}
\Arc(45,17)(40,200,270)
\Arc(45,17)(40,270,340)
\Vertex(5,5){2}
\Vertex(85,5){2}
\Vertex(48,33){2}
%\SetWidth{1.0}
%\Vertex(5,5){2}
\Line(5,5)(-5,5)
\Line(85,5)(95,5)
%\Vertex(40,5){2}
%\Vertex(40,15){2}
%\Vertex(40,-15){2}
%\Text(67,13)[b]{$\scriptstyle k$}
%\Text(65,10)[b]{$m$}
%\Text(33,3)[t]{$\scriptstyle \frac{M^2}{s(1-s)}$}
%\Text(33,1)[t]{$\scriptstyle M^2/[s(1-s)]$}
%\Text(65,-25)[t]{$\alpha$}
%\Text(33,7)[t]{$2$}
\Text(25,-20)[t]{$2$}
\Text(-3,-5)[b]{$\to$}
\Text(-3,-12)[b]{$q$}
%\Text(-3,-5)[b]{$p$}
\end{axopicture}
  }} \hspace{0.5cm} - \frac{m^2}{q^2} \,  
\hspace{3mm}
%\nonumber \\ &&
%\nonumber \\ &&
%\nonumber \\ &&
%\nonumber \\ &&
  \raisebox{1mm}{{
    %\begin{picture}(90,30)(0,4)
    \begin{axopicture}(90,10)(0,4)
%  \SetWidth{2.0}
  \SetWidth{0.5}
%\CArc(5,5)(80,20,160)
%\CArc(45,5)(40,0,180)
\Arc(45,-7)(40,20,160)
  %\Line[arrow](5,5)(40,5)
%\Line[arrow](40,5)(85,5)
\SetWidth{1.5}
\Arc(5,-35)(40,20,85)
\SetWidth{0.5}
\Arc(45,17)(40,200,270)
\SetWidth{0.5}
\Arc(45,17)(40,270,340)
\SetWidth{0.5}
\Vertex(5,5){2}
\Vertex(85,5){2}
\Vertex(43,-23){2}
%\SetWidth{1.0}
%\Vertex(5,5){2}
\Line(5,5)(-5,5)
\Line(85,5)(95,5)
%\Vertex(40,5){2}
%\Vertex(40,15){2}
%\Vertex(40,-15){2}
%\Text(65,-10)[b]{$n$}
%\Text(65,10)[b]{$m$}
%\Text(33,3)[t]{$\scriptstyle \frac{M^2}{s(1-s)}$}
%\Text(33,1)[t]{$\scriptstyle M^2/[s(1-s)]$}
%\Text(65,-25)[t]{$\alpha-1$}
%\Text(33,7)[t]{$2$}
\Text(25,-20)[t]{$2$}
\Text(-3,-5)[b]{$\to$}
\Text(-3,-12)[b]{$q$}
%\Text(-3,-5)[b]{$p$}
\end{axopicture}
  }}
\hspace{3mm} \, .
\label{eq4}
\ee
\vskip 0.5cm

Because
\be
\frac{1}{(q^2+m^2)^2} = - \frac{d}{dm^2} \, \frac{1}{(q^2+m^2)^2}
\label{diff}
\ee
Eq. (\ref{eq4}) can be rewritten in the form
\be
\left[(d-4)  \, \left(1-\frac{2m^2}{q^2}\right) \, - 2m^2 \, \left(1-\frac{m^2}{q^2}\right) \, \frac{d}{dm^2} \, \right] I_1(q^2,m^2) = 2 \, J_1(q^2,m^2)
\, ,
\label{eq4a}
\ee
i.e. the first order DE
\footnote{Hereafter we consider only first order DEs. The consideration of the high order DEs can be found in Section 7 of the review
\cite{Blumlein:2018cms}. See also the recent papers \cite{Klemm:2019dbm}.}
%differential equation
for the original diagram with the inhomogeneous term $J_1(q^2,m^2)$ containing only simpler diagrams, 
i.e. those obtained from the original expression by canceling one of the propagators, see eq. (\ref{eq4}).

The first diagram in the inhomogeneous term $J_1(q^2,m^2)$ is independent of mass and can therefore be easily calculated as
a product of the $\Gamma$-functions, see Eq. (\ref{Lp}) below,
\vskip 0.5cm
\be
\hspace{3mm}
  \raisebox{1mm}{{
    %\begin{picture}(90,30)(0,4) 
    \begin{axopicture}(90,10)(0,4)
%  \SetWidth{2.0}
  \SetWidth{0.5}
\CArc(25,5)(20,0,180)
\CArc(25,5)(20,180,360)
%  \Arc(20,-7)(20,20,160)
%  \Arc(20,17)(20,200,340)
%\Line[arrow](5,5)(40,5)
%\Line[arrow](40,5)(85,5)
 \Arc(65,5)(20,0,180)
  \Arc(65,5)(20,180,360)
  \Vertex(5,5){2}
  \Vertex(45,5){2}
\Vertex(85,5){2}
%\SetWidth{1.0}
%\Vertex(5,5){2}
\Line(5,5)(-5,5)
\Line(85,5)(95,5)
%\Vertex(40,5){2}
%\Vertex(40,15){2}
%\Vertex(40,-15){2}
%\Text(65,-10)[b]{$n$}
%\Text(65,10)[b]{$m$}
%\Text(33,3)[t]{$\scriptstyle \frac{M^2}{s(1-s)}$}
%\Text(33,1)[t]{$\scriptstyle M^2/[s(1-s)]$}
%\Text(45,27)[t]{$\alpha_1$}
%\Text(65,-20)[t]{$\alpha$}
\Text(25,-20)[t]{$2$}
\Text(-3,-5)[b]{$\to$}
\Text(-3,-12)[b]{$q$}
%\Text(-3,-5)[b]{$p$}
\end{axopicture}
}}
  \hspace{3mm}
= L_{2,1}(q^2) L_{1,1}(q^2) =  \frac{1}{(4\pi)^{d}} \, \frac{A(2,1)A(1,1)}{q^{2(5-d)}} \, .
\label{J11}
\ee
\vskip 1cm
\noindent
where $A(\alpha_1,\alpha_2)$ is given in eq. (\ref{R}) below.

Using IBP relations, for the remaining two diagrams in the inhomogeneous $J_1(q^2,m^2)$ term diagrams, one can obtain
similar equations with inhomogeneous terms containing only even simpler diagrams, i.e. those obtained from the original by canceling two propagators.
These results are given
%can be found
in the Appendix.

%%%%%%%%%%%%%%%%% NEW %%%%%%%%%%%
\section{Calculation of massive Feynman integrals}

%Feynman integrals with massive propagators are significantly more complex objects compared to the
%massless case. The basic rules for calculating such diagrams are discussed in Section 2, which are
%supplemented by new ones containing directly massive propagators. These additional rules are discussed
%in the next subsection.
%\subsection{Basic formulas}

Let us briefly consider the rules for calculating diagrams having
%with
the massive propagators.\\

{\bf 1.}~~ The massless propagator and the propagator with mass $m$ will be represented as
\be
\frac{1}{q^{2\alpha}}
%\equiv  \frac{1}{q^{2\alpha}}
= \hspace{3mm} \raisebox{1mm}{{
%\begin{picture}(70,30)(0,4)
\begin{axopicture}(70,30)(0,4)
%    \SetWidth{2.0}
  \SetWidth{0.5}
\Line(5,5)(65,5)
\SetWidth{0.5}
%\Photon(5,5)(65,5){5}-{7}
\Vertex(5,5){2}
\SetWidth{1.0}
\Vertex(65,5){2}
\Line(5,5)(-5,5)
\Line(65,5)(75,5)
%\Text(33,7)[b]{$\scriptstyle a+b-d/2$}
%\Text(33,7)[b]{}
%\Text(33,10)[b]{$M$}
%\Text(33,3)[t]{$\scriptstyle \frac{M^2}{s(1-s)}$}
%\Text(33,1)[t]{$\scriptstyle M^2/[s(1-s)]$}
%\Text(33,1)[t]{$\alpha$}
\Text(33,-1)[t]{$\alpha$}
\Text(-3,-5)[b]{$\to$}
\Text(-3,-12)[b]{$q$}
%\Text(-3,-5)[b]{$p$}
\end{axopicture}
}}
\hspace{3mm} \, ,
~~~
%D_{\alpha}(q) =
\frac{1}{(q^2+m^2)^{\alpha}}
%\equiv  \frac{1}{q^{2\alpha}}
= \hspace{3mm} \raisebox{1mm}{{
%\begin{picture}(70,30)(0,4)
\begin{axopicture}(70,30)(0,4)
%    \SetWidth{2.0}
  \SetWidth{1.5}
\Line(5,5)(65,5)
\SetWidth{0.5}
%\Photon(5,5)(65,5){5}-{7}
\Vertex(5,5){2}
\SetWidth{1.0}
\Vertex(65,5){2}
\Line(5,5)(-5,5)
\Line(65,5)(75,5)
%\Text(33,7)[b]{$\scriptstyle a+b-d/2$}
%\Text(33,7)[b]{}
\Text(33,10)[b]{$m$}
%\Text(33,3)[t]{$\scriptstyle \frac{M^2}{s(1-s)}$}
%\Text(33,1)[t]{$\scriptstyle M^2/[s(1-s)]$}
%\Text(33,1)[t]{$\alpha$}
\Text(33,-1)[t]{$\alpha$}
\Text(-3,-5)[b]{$\to$}
\Text(-3,-12)[b]{$q$}
%\Text(-3,-5)[b]{$p$}
\end{axopicture}
}}
\hspace{3mm}
,~~
%\nonumber
\label{DefM}
\ee
\vskip 0.5cm
\noindent
where the symbol $m$ will be omitted in the single-mass case (as in the case of $I_1(q^2,m^2)$ in eq. (\ref{I1})).\\

 {\bf 2.}~~The massive one-loop tadpole $T_{\alpha_1,\alpha_2}(m^2)$ and the massless loop $L_{\alpha_1,\alpha_2}(q^2)$ can be calculated exactly
as combinations of the $\Gamma$-functions:
\vskip 0.5cm
\bea
&& T_{\alpha_1,\alpha_2}(m^2)=
\int \frac{Dk }{k^{2\alpha_1}(k^2+m^2)^{\alpha_2}}
= \hspace{3mm}
\raisebox{1mm}{{
    %\begin{picture}(90,30)(0,4)
    \begin{axopicture}(90,10)(0,4)
%  \SetWidth{2.0}
  \SetWidth{0.5}
%\CArc(5,5)(80,20,160)
%\CArc(45,5)(40,0,180)
\Arc(45,-7)(40,20,160)
  %\Line[arrow](5,5)(40,5)
%\Line[arrow](40,5)(85,5)
\SetWidth{1.5}
\Arc(45,17)(40,200,340)
\SetWidth{0.5}
\Vertex(5,5){2}
\Vertex(85,5){2}
%\SetWidth{1.0}
%\Vertex(5,5){2}
\Line(5,5)(-5,0)
\Line(5,5)(-5,10)
%\Line(85,5)(95,5)
%\Vertex(40,5){2}
%\Vertex(40,15){2}
%\Vertex(40,-15){2}
\Text(45,-16)[b]{$\alpha_2$}
%\Text(65,10)[b]{$m$}
%\Text(33,3)[t]{$\scriptstyle \frac{M^2}{s(1-s)}$}
%\Text(33,1)[t]{$\scriptstyle M^2/[s(1-s)]$}
\Text(45,27)[t]{$\alpha_1$}
%\Text(45,-29)[t]{$\alpha_2$}
%\Text(-3,-5)[b]{$\to$}
%\Text(-3,-12)[b]{$q$}
%\Text(-3,-5)[b]{$p$}
\end{axopicture}
}}
\hspace{3mm}
=
%\frac{1}{(4\pi)^{d/2}} \,
\frac{R(\alpha_1,\alpha_2)}{m^{2(\alpha_1+\alpha_2-d/2)}} \, ,
\label{Tp}\\
&&\nonumber \\ && \nonumber \\
&& \nonumber \\
&&L_{\alpha_1,\alpha_2}(q^2)= \int \frac{Dk}{(q-k)^{2\alpha_1}k^{2\alpha_2}} 
= \hspace{3mm}
\raisebox{1mm}{{
    %\begin{picture}(90,30)(0,4)
    \begin{axopicture}(90,10)(0,4)
%  \SetWidth{2.0}
  \SetWidth{0.5}
%\CArc(5,5)(80,20,160)
%\CArc(45,5)(40,0,180)
\Arc(45,-7)(40,20,160)
  %\Line[arrow](5,5)(40,5)
%\Line[arrow](40,5)(85,5)
\Arc(45,17)(40,200,340)
\Vertex(5,5){2}
\Vertex(85,5){2}
%\SetWidth{1.0}
%\Vertex(5,5){2}
\Line(5,5)(-5,5)
\Line(85,5)(95,5)
%\Vertex(40,5){2}
%\Vertex(40,15){2}
%\Vertex(40,-15){2}
\Text(45,-16)[b]{$\alpha_2$}
%\Text(65,10)[b]{$m$}
%\Text(33,3)[t]{$\scriptstyle \frac{M^2}{s(1-s)}$}
%\Text(33,1)[t]{$\scriptstyle M^2/[s(1-s)]$}
\Text(45,27)[t]{$\alpha_1$}
%\Text(45,-29)[t]{$\alpha_2$}
\Text(-3,-5)[b]{$\to$}
\Text(-3,-12)[b]{$q$}
%\Text(-3,-5)[b]{$p$}
\end{axopicture}
}}
\hspace{3mm}
=
%\frac{1}{(4\pi)^{d/2}} \,
\frac{A(\alpha_1,\alpha_2)}{q^{2(\alpha_1+\alpha_2-d/2)}} \, ,
\label{Lp}
\eea
\vskip 0.5cm
where
\bea
&&A(\alpha_1,\alpha_2) = \frac{a(\alpha_1)a(\alpha_2)}{a(\alpha_1+\alpha_2-d/2)},~~ a(\alpha)=\frac{\Gamma(\tilde{\alpha})}{\Gamma(\alpha)},~~
\tilde{\alpha}=\frac{d}{2}-\alpha \, ,
\label{A}\\
&&R(\alpha_1,\alpha_2) = \frac{\Gamma(d/2-\alpha_1)\Gamma(\alpha_1+\alpha_2-d/2)}{\Gamma(d/2)\Gamma(\alpha_2)}
 \, 
\label{R}
\eea
and
\be
Dk = \frac{d^dk}{\pi^{d/2}} = (4\pi)^{d/2} \, D_Ek,~~ D_Ek = \frac{d^dk}{(2\pi)^d} \, .
 \label{Measure}
\ee
Here $D_Ek$ is the usual  Euclidean measure in
%integration
 $d=4-2\ep$ space.\\
%Euclidean measure.\\

{\bf 3.}~~ A simple loop of two massive propagators with masses $m_1$ and $m_2$ can be represented
as hypergeometric function, which can be calculated in a general form, for example, by Feynman-parameter
method, see \cite{Ryder}.
It is very convenient, using this approach to represent the loop as an integral of a propagator
with the ``effective mass'' $\mu$
\cite{Kotikov:1990kg,Fleischer:1999hp,Fleischer:1997bw,FleKoVe,Kotikov:2020ccc,Kniehl:2005bc,Kniehl:2005yc}:
\bea
&&
%(4\pi)^{d/2} \times
\int \frac{Dk }{[(q-k)^2+m_1^2]^{\alpha_1}[k^2+m_2^2]^{\alpha_2}} \,
%\nonumber \\&&
= \,
%\frac{1}{(4\pi)^{d/2}} \,
\frac{\Gamma(\alpha_1+\alpha_2-d/2)}{\Gamma(\alpha_1)\Gamma(\alpha_2)} \,
\nonumber \\&&\times
\int_0^1 \, \frac{ds \, s^{\alpha_1-1} \, (1-s)^{\alpha_2-1} }{[s(1-s)q^2+m_1^2s + m_2^2(1-s)]^{\alpha_1+\alpha_2-d/2}}
%\nonumber \\&&
\, = \,
%\frac{1}{(4\pi)^{d/2}} \,
\frac{\Gamma(\alpha_1+\alpha_2-d/2)}{\Gamma(\alpha_1)\Gamma(\alpha_2)} \,
\nonumber \\&&\times
\int_0^1 \, \frac{ds}{s^{1-\tilde{\alpha}_2} \, (1-s)^{1-\tilde{\alpha}_1} } \,
\frac{1}{[q^2+\mu^2]^{\alpha_1+\alpha_2-d/2}},~~
\left(\mu^2 = \frac{m_1^2}{1-s} +  \frac{m_2^2}{s}\right) \, .
  %  \label{Loop}
\nonumber
\eea

It is useful
%convenient
to rewrite the equation graphically as
\vskip 0.5cm
\be
%&&\int \frac{Dk}{(4\pi)^D} \, D_{\alpha_1}(q_1-k)D^{n}_{\alpha_2}(k-q_2) = A^{0,n}(\alpha_1,\alpha_2) \,
%D^{n}_{\alpha_1 + \alpha_2-D/2}(q_1-q_2) + ... \, , \nonumber \\
%%label{loop}
%%eea
%&& %\hspace{-1cm} 
%\mbox{or graphically} \nonumber \\ &&
%\nonumber \\ 
%\bea&&
%\hspace{-5mm}
\raisebox{1mm}{{
    %\begin{picture}(90,30)(0,4)
    \begin{axopicture}(90,10)(0,4)
%  \SetWidth{2.0}
  \SetWidth{1.5}
%\CArc(5,5)(80,20,160)
%\CArc(45,5)(40,0,180)
\Arc(45,-7)(40,20,160)
  %\Line[arrow](5,5)(40,5)
%\Line[arrow](40,5)(85,5)
\Arc(45,17)(40,200,340)
 \SetWidth{1.0}
\Vertex(5,5){2}
\Vertex(85,5){2}
%\SetWidth{1.0}
%\Vertex(5,5){2}
\Line(5,5)(-5,5)
\Line(85,5)(95,5)
%\Vertex(40,5){2}
%\Vertex(40,15){2}
%\Vertex(40,-15){2}
\Text(45,-16)[b]{$m_2$}
\Text(45,40)[b]{$m_1$}
%\Text(33,3)[t]{$\scriptstyle \frac{M^2}{s(1-s)}$}
%\Text(33,1)[t]{$\scriptstyle M^2/[s(1-s)]$}
\Text(45,27)[t]{$\alpha_1$}
\Text(45,-29)[t]{$\alpha_2$}
\Text(-3,-5)[b]{$\to$}
\Text(-3,-12)[b]{$q$}
%\Text(-3,-5)[b]{$p$}
\end{axopicture}
}}
\hspace{3mm}
=
%\frac{1}{(4\pi)^{d/2}} \,
\frac{\Gamma(\alpha_1+\alpha_2-d/2)}{\Gamma(\alpha_1)\Gamma(\alpha_2)} \,
\int_0^1 \, \frac{ds}{s^{1-\tilde{\alpha}_2} \, (1-s)^{1-\tilde{\alpha}_1} } \,
%\frac{1}{(4\pi)^{d/2}} \, A^{0,n}(\alpha_1,\alpha_2) \,
\hspace{3mm} \raisebox{1mm}{{
\begin{picture}(70,30)(0,4)
%  \SetWidth{2.0}
  \SetWidth{2.0}
\Line(5,5)(65,5)
\SetWidth{1.0}
\Vertex(5,5){2}
\Vertex(65,5){2}
\Line(5,5)(-5,5)
\Line(65,5)(75,5)
\Text(33,10)[b]{$\mu$}
%\Text(33,7)[b]{}
%\Text(33,3)[t]{$\scriptstyle \frac{M^2}{s(1-s)}$}
%\Text(33,1)[t]{$\scriptstyle M^2/[s(1-s)]$}
\Text(33,-1)[t]{$\scriptstyle \alpha_1+\alpha_2-d/2$}
\Text(-3,-5)[b]{$\to$}
\Text(-3,-12)[b]{$q$}
%\Text(-3,-5)[b]{$q$}
\end{picture}
}}
\hspace{3mm}
% + ... \, ,
\, .
%\nonumber 
\label{loopM}
 \ee

\vskip 1.2cm

The rule is very convenient in the cases $m_2=0$ and $m_1=m_2$, where the variable $\mu$ is equal to
%corresponding
$\mu^2=m_1^2/s$ and $\mu^2=m_1^2/s(1-s)$, respectively. Such simple forms of $\mu$ provide the possibility to use directly
an inverse-mass expansion without applying the Mellin-Barnes representation, which is essentially more complicated procedure.\\

{\bf 4.}~~ For any triangle with indices
$\alpha_i$ ($i=1,2,3$) and masses $m_i$  there is the following relation, which is based on
%Triangle (
integration by parts procedure \cite{Chetyrkin:1981qh,Kotikov:1990kg,Kotikov:1991hm}
%):
%\bea
%&&(D-2\alpha_1-\alpha_2-\alpha_3+n+m+k)
%\int \frac{Dk \, \prod_{i=1}^{n} (q_1-k)^{\mu_i} \, \prod_{j=1}^{m} (q_1-k)^{\nu_j} \, \prod_{l=1}^{k} (q_3-k)^{\mu_l}}{(q_1-k)^{2\alpha_1} (q_2-k)^{2\alpha_2} (q_3-k)^{2\alpha_3}}
%\nonumber \\ &&
%=\alpha_2\int \frac{Dk \, \prod_{l=1}^{k} (q_3-k)^{\mu_l}}{ (q_3-k)^{2\alpha_3}} \Biggl[ \frac{\prod_{i=1}^{n} (q_1-k)^{\mu_i} \, \prod_{j=1}^{m} (q_1-k)^{\nu_j}}{
%  (q_1-k)^{2(\alpha_1-1)} (q_2-k)^{2(\alpha_2+1)}} 
%  -(q_2-q_1)^2 \times  \frac{\prod_{i=1}^{n} (q_1-k)^{\mu_i} \, \prod_{j=1}^{m} (q_1-k)^{\nu_j}}{
%  (q_1-k)^{2\alpha_1} (q_1-k)^{2(\alpha_2+1)}}  \nonumber \\
%&&  +
%  m (q_2-q_1)^{\mu_m} \times \frac{\prod_{i=1}^{n} (q_1-k)^{\mu_i} \, \prod_{j=1}^{m-1} (q_1-k)^{\nu_j}}{
%    (q_1-k)^{2(\alpha_1-1)} (q_2-k)^{2(\alpha_2+1)}} \Biggr]
%\nonumber \\&&
%+ \alpha_3
%\int \frac{Dk}{(4\pi)^D}
%\, \biggl[\alpha_2 \leftrightarrow \alpha_3, m \leftrightarrow k \biggr] \, ,
%\label{TreIBP}
%\nonumber
%\eea
%
%\bea
%&&(D-2\alpha_1-\alpha_2-\alpha_3+n+m+k)
%\int \frac{Dk}{(4\pi)^D} \, D_{\alpha_1}^n(q_1-k)D^{m}_{\alpha_2}(q_2-k) D^{k}_{\alpha_3}(q_3-k)
%\nonumber \\ &&
%=\alpha_2\int \frac{Dk}{(4\pi)^D} \, \biggl[ D_{\alpha_1-1}^n(q_1-k)D^{m}_{\alpha_2+1}(q_2-k)
%  -(q_2-q_1)^2 \times  D_{\alpha_1-1}^n(q_1-k)D^{m}_{\alpha_2+1}(q_2-k) \nonumber \\
%&&  +
%  m (q_2-q_1)^{\mu_m} \times  D_{\alpha_1-1}^n(q_1-k)D^{m-1}_{\alpha_2+1}(q_2-k) \Biggr] D^{k}_{\alpha_3}(q_3-k) \nonumber \\
%&&+ \alpha_3 \int \frac{Dk}{(4\pi)^D} \, \biggl[\alpha_2 \leftrightarrow \alpha_3, m \leftrightarrow k \biggr] \, ,
%\label{TreIBP}
%\nonumber
%\eea
%or graphically
\vskip 1cm

\bea
&& (d-2\alpha_1-\alpha_2-\alpha_3) \hspace{0.5cm}
\raisebox{1mm}{{
    %\begin{picture}(90,30)(0,4)
    \begin{axopicture}(90,10)(0,4)
%  \SetWidth{2.0}
  \SetWidth{1.5}
%\CArc(5,5)(80,20,160)
%\CArc(45,5)(40,0,180)
%\Arc(45,-7)(40,20,160)
\Line(5,5)(45,45)
\Line(5,5)(85,5)
\Line(45,45)(85,5)
\SetWidth{0.5}
%\Arc[arrow](45,17)(40,200,340)
\Vertex(5,5){2}
\Vertex(85,5){2}
\Vertex(45,45){2}
%\SetWidth{1.0}
%\Vertex(5,5){2}
\Line(5,5)(-5,5)
\Line(85,5)(90,5)
\Line(45,45)(55,60)
%\Vertex(40,5){2}
%\Vertex(40,15){2}
%\Vertex(40,-15){2}
\Text(45,10)[b]{$\scriptstyle m_1$}
\Text(20,30)[b]{$\scriptstyle m_2$}
\Text(70,30)[b]{$\scriptstyle m_3$}
%\Text(33,3)[t]{$\scriptstyle \frac{M^2}{s(1-s)}$}
%\Text(33,1)[t]{$\scriptstyle M^2/[s(1-s)]$}
\Text(35,25)[t]{$\scriptstyle \alpha_2$}
\Text(45,-2)[t]{$\scriptstyle \alpha_1$}
\Text(55,25)[t]{$\scriptstyle \alpha_3$}
\Text(-3,-5)[b]{$\to$}
\Text(-3,-12)[b]{$\scriptstyle q_2-q_1$}
\Text(93,-5)[b]{$\to$}
\Text(93,-12)[b]{$\scriptstyle q_1-q_3$}
\Text(70,50)[b]{$\to$}
\Text(70,40)[b]{$\scriptstyle q_3-q_2$}
%\Text(-3,-5)[b]{$p$}
\end{axopicture}
}}
\hspace{3mm} \nonumber \\
%\vspace{1cm} 
&&\nonumber \\
&&\nonumber \\
&&\nonumber \\
&&= \alpha_2 \biggl[ \, \hspace{0.5cm} \,
  \raisebox{1mm}{{
    %\begin{picture}(90,30)(0,4)
    \begin{axopicture}(90,10)(0,4)
%  \SetWidth{2.0}
  \SetWidth{1.5}
%\CArc(5,5)(80,20,160)
%\CArc(45,5)(40,0,180)
%\Arc(45,-7)(40,20,160)
  \Line(5,5)(45,45)
\Line(5,5)(85,5)
\Line(45,45)(85,5)
\SetWidth{0.5}
%\Arc[arrow](45,17)(40,200,340)
\Vertex(5,5){2}
\Vertex(85,5){2}
\Vertex(45,45){2}
%\SetWidth{1.0}
%\Vertex(5,5){2}
\Line(5,5)(-5,5)
\Line(85,5)(90,5)
\Line(45,45)(55,60)
%\Vertex(40,5){2}
%\Vertex(40,15){2}
%\Vertex(40,-15){2}
\Text(45,10)[b]{$\scriptstyle m_1$}
\Text(20,30)[b]{$\scriptstyle m_2$}
\Text(70,30)[b]{$\scriptstyle m_3$}
%\Text(33,3)[t]{$\scriptstyle \frac{M^2}{s(1-s)}$}
%\Text(33,1)[t]{$\scriptstyle M^2/[s(1-s)]$}
\Text(35,25)[t]{$\scriptstyle \alpha_2+1$}
\Text(45,-2)[t]{$\scriptstyle \alpha_1-1$}
\Text(55,25)[t]{$\scriptstyle \alpha_3$}
\Text(-3,-5)[b]{$\to$}
\Text(-3,-12)[b]{$\scriptstyle q_2-q_1$}
\Text(93,-5)[b]{$\to$}
\Text(93,-12)[b]{$\scriptstyle q_1-q_3$}
\Text(70,50)[b]{$\to$}
\Text(70,40)[b]{$\scriptstyle q_3-q_2$}
%\Text(-3,-5)[b]{$p$}
\end{axopicture}
}}
\hspace{3mm}
- \biggl[(q_2-q_1)^2 +m_1^2 +m_2^2 \biggr] \times
  \raisebox{1mm}{{
    %\begin{picture}(90,30)(0,4)
    \begin{axopicture}(90,10)(0,4)
%  \SetWidth{2.0}
  \SetWidth{1.5}
%\CArc(5,5)(80,20,160)
%\CArc(45,5)(40,0,180)
  %\Arc(45,-7)(40,20,160)
\Line(5,5)(45,45)
\Line(5,5)(85,5)
\Line(45,45)(85,5)
 \SetWidth{0.5}
%\Arc[arrow](45,17)(40,200,340)
\Vertex(5,5){2}
\Vertex(85,5){2}
\Vertex(45,45){2}
%\SetWidth{1.0}
%\Vertex(5,5){2}
\Line(5,5)(-5,5)
\Line(85,5)(90,5)
\Line(45,45)(55,60)
%\Vertex(40,5){2}
%\Vertex(40,15){2}
%\Vertex(40,-15){2}
\Text(45,10)[b]{$\scriptstyle m_1$}
\Text(20,30)[b]{$\scriptstyle m_2$}
\Text(70,30)[b]{$\scriptstyle m_3$}
%\Text(33,3)[t]{$\scriptstyle \frac{M^2}{s(1-s)}$}
%\Text(33,1)[t]{$\scriptstyle M^2/[s(1-s)]$}
\Text(35,25)[t]{$\scriptstyle \alpha_2+1$}
\Text(45,-2)[t]{$\scriptstyle \alpha_1$}
\Text(55,25)[t]{$\scriptstyle \alpha_3$}
\Text(-3,-5)[b]{$\to$}
\Text(-3,-12)[b]{$\scriptstyle q_2-q_1$}
\Text(93,-5)[b]{$\to$}
\Text(93,-12)[b]{$\scriptstyle q_1-q_3$}
\Text(70,50)[b]{$\to$}
\Text(70,40)[b]{$\scriptstyle q_3-q_2$}
%\Text(-3,-5)[b]{$p$}
\end{axopicture}
}} 
\hspace{3mm}
\Biggr]
\nonumber \\
%\vspace{1cm} 
&&\nonumber \\
&&\nonumber \\
&&\nonumber \\
&&
%+ \Biggr]
+ \alpha_3
%\int \frac{Dk}{(4\pi)^D}
\, \biggl[\alpha_2 \leftrightarrow \alpha_3, m_2 \leftrightarrow m_3 \biggr] 
%m (q_2-q_1)^{\mu_m}
-2m_1^2 \alpha_1 
\times
\raisebox{1mm}{{
    %\begin{picture}(90,30)(0,4)
    \begin{axopicture}(90,10)(0,4)
%  \SetWidth{2.0}
  \SetWidth{1.5}
%\CArc(5,5)(80,20,160)
%\CArc(45,5)(40,0,180)
%\Arc(45,-7)(40,20,160)
\Line(5,5)(45,45)
\Line(5,5)(85,5)
\Line(45,45)(85,5)
 \SetWidth{0.5}
%\Arc[arrow](45,17)(40,200,340)
\Vertex(5,5){2}
\Vertex(85,5){2}
\Vertex(45,45){2}
%\SetWidth{1.0}
%\Vertex(5,5){2}
\Line(5,5)(-5,5)
\Line(85,5)(90,5)
\Line(45,45)(55,60)
%\Vertex(40,5){2}
%\Vertex(40,15){2}
%\Vertex(40,-15){2}
\Text(45,10)[b]{$\scriptstyle m_1$}
\Text(20,30)[b]{$\scriptstyle m_2$}
\Text(70,30)[b]{$\scriptstyle m_3$}
%\Text(33,3)[t]{$\scriptstyle \frac{M^2}{s(1-s)}$}
%\Text(33,1)[t]{$\scriptstyle M^2/[s(1-s)]$}
\Text(35,25)[t]{$\scriptstyle \alpha_2$}
\Text(45,-2)[t]{$\scriptstyle \alpha_1+1$}
\Text(55,25)[t]{$\scriptstyle \alpha_3$}
\Text(-3,-5)[b]{$\to$}
\Text(-3,-12)[b]{$\scriptstyle q_2-q_1$}
\Text(93,-5)[b]{$\to$}
\Text(93,-12)[b]{$\scriptstyle q_1-q_3$}
\Text(70,50)[b]{$\to$}
\Text(70,40)[b]{$\scriptstyle q_3-q_2$}
%\Text(-3,-5)[b]{$p$}
\end{axopicture}
}}
\hspace{5mm}
\, .
%\nonumber
\label{TreIBPM}
\eea

\vskip 0.5cm

Eq. (\ref{TreIBPM}) can been obtained by introducing the factor $(\partial/\partial k_{\mu}) \, (k-q_1)^{\mu}$ to the subintegral
expression of the triangle, shown below as $[...]$,
and using the integration by parts procedure as follows:
\bea
&& d \int Dk \, \bigl[ ...\bigr] = \int Dk \, \left(\frac{\partial}{\partial k_{\mu}} \, (k-q_1)^{\mu}\right) \,  \bigl[ ...\bigr]      =
\int Dk \,  \frac{\partial}{\partial k_{\mu}} \, \left((k-q_1)^{\mu} \,  \bigl[ ...\bigr] \right) \nonumber \\&&
- \int Dk \, (k-q_1)^{\mu} \,
\frac{\partial}{\partial k_{\mu}} \, \left( \bigl[ ...\bigr]\right)
\label{IBPpro}
\eea
The first term in the r.h.s. becomes to be zero because it can be represented as a surface integral on the infinite surface.
Evaluating the second  term in the r.h.s. we  reproduce Eq. (\ref{TreIBPM}).
Note that the equation (\ref{IBPpro}) can also be applied to the $n$-point subgraph, see, for example, \cite{Kotikov:1991pm}. 

As it is possible to see from Eqs. (\ref{TreIBPM}) and (\ref{IBPpro}) the line with the index $\alpha_1$ is distinguished. The
contributions of the other lines are the same. So, we will denote below the line with the index $\alpha_1$ as a ``distinguished line''.
It is clear that a various choices of the distinguished line produce
different types of the IBP relations.

%Using equation (\ref{TreIBP}) allows you to change the indices of the line diagrams by an integer. You can also change line indices using the point group of transformations [3]. The elements of the group are:
%a) the transition to impulse presentation,
%b) conformal inversion transformation $p \to p' =p/p^2$, c) a special series of transformations that makes it possible to make one of the vertices unique, and then apply relation (3) to it.

\subsection{Basic massive two-loop integrals}

%Consider the action of the transformation group on the example of a characteristic two-loop diagram depicted in Fig. 1.
%The table of transformations is given in the appendix. The notation used there is given in [3].
%The general topology of the two-loop two-point diagram, which cannot be expressed as a combination of loops and chanins is shown on Fig.4.
%\vskip 2cm
\vskip 0.5cm
\begin{figure}[htbp]
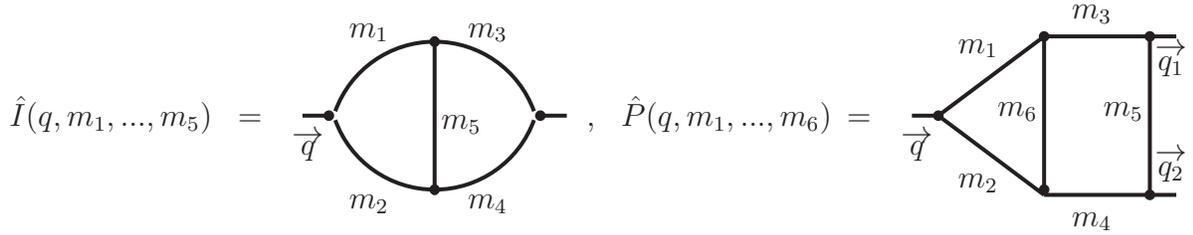

\centerline{
$\hat{I}(q,m_1,...,m_5)$ \, =
 \hspace{3mm}
%\frac{2^{n-m} \Gamma(n-m+\alpha)}{(n-m)!\Gamma(\alpha)} \, \hspace{3mm}
\raisebox{1mm}{{
    %\begin{picture}(90,30)(0,4)
    \begin{axopicture}(90,10)(0,4)
%  \SetWidth{2.0}
  \SetWidth{1.5}
%\CArc(5,5)(80,20,160)
%\CArc(45,5)(40,0,180)
%\Arc(45,-7)(40,20,160)
\Arc(45,-7)(40,90,160)
%\SetWidth{0.7}
\Arc(45,-7)(40,20,90)
%\Arc[arrow](45,-7)(40,90,160)
\Arc(45,17)(40,200,270)
\Line(45,-25)(45,35)
%\Line[arrow](40,5)(85,5)
%\Arc(45,17)(40,200,270)
\Arc(45,17)(40,270,340)
%\Arc(45,17)(40,200,340)
\Vertex(45,-23){2}
\Vertex(45,33){2}
\Vertex(5,5){2}
\Vertex(85,5){2}
%\SetWidth{1.0}
%\Vertex(5,5){2}
\Line(5,5)(-5,5)
\Line(85,5)(95,5)
%Line(45,-25)(45,-35)
%Line(85,5)(95,10)
%\Vertex(40,5){2}
%\Vertex(40,15){2}
%\Vertex(40,-15){2}
\Text(20,33)[b]{$m_1$}
\Text(20,-25)[t]{$m_2$}
\Text(65,33)[b]{$m_3$}
\Text(65,-25)[t]{$m_4$}
\Text(55,5)[t]{$m_5$}
%\Text(33,3)[t]{$\scriptstyle \frac{M^2}{s(1-s)}$}
%\Text(33,1)[t]{$\scriptstyle M^2/[s(1-s)]$}
%\Text(75,-22)[t]{$\alpha$}
%\Text(75,-22)[t]{$\scriptstyle \alpha$}
\Text(-3,-5)[b]{$\to$}
\Text(-3,-12)[b]{$q$}
%\Text(93,-5)[b]{$\to$}
%\Text(93,-12)[b]{$q$}
%\Text(93,20)[b]{$\to$}
%\Text(93,12)[b]{$p$}
%\Text(52,-30)[b]{$\to$}
%\Text(52,-38)[b]{$p$}
%\Text(-3,-5)[b]{$p$}
\end{axopicture}
}}
\hspace{3mm},
\,
$\hat{P}(q,m_1,...,m_6)$\, =
 \hspace{3mm}
%\frac{2^{n-m} \Gamma(n-m+\alpha)}{(n-m)!\Gamma(\alpha)} \, \hspace{3mm}
\raisebox{1mm}{{
    %\begin{picture}(90,30)(0,4)
    \begin{axopicture}(90,10)(0,4)
%  \SetWidth{2.0}
  \SetWidth{1.5}
%\CArc(5,5)(80,20,160)
%\CArc(45,5)(40,0,180)
%\Arc(45,-7)(40,20,160)
%\Arc(45,-7)(40,20,90)
%%\Arc[arrow](45,-7)(40,90,160)
%\Arc(45,-7)(40,90,160)
  \Line(5,5)(45,35)
  \Line(5,5)(45,-25)
\Line(45,-25)(45,35)
%\SetWidth{0.7}
\Line(45,35)(85,35)
\Line(45,-25)(85,-25)
\Line(85,35)(85,-25)
%\Arc(45,17)(40,200,270)
%\Arc[arrow](45,17)(40,270,340)
%\Arc(45,17)(40,200,340)
\Vertex(45,35){2}
\Vertex(45,-23){2}
\Vertex(5,5){2}
\Vertex(85,35){2}
\Vertex(85,-25){2}
%\SetWidth{1.0}
%\Vertex(5,5){2}
\Line(5,5)(-5,5)
\Line(85,35)(95,35)
\Line(85,-25)(95,-25)
%Line(45,-25)(45,-35)
%Line(85,5)(95,10)
%\Vertex(40,5){2}
%\Vertex(40,15){2}
%\Vertex(40,-15){2}
\Text(20,27)[b]{$m_1$}
\Text(20,-17)[t]{$m_2$}
\Text(63,40)[b]{$m_3$}
\Text(63,-35)[7]{$m_4$}
\Text(75,3)[b]{$m_5$}
\Text(35,3)[b]{$m_6$}
%\Text(75,-22)[t]{$\alpha$}
%\Text(75,-22)[t]{$\scriptstyle \alpha$}
\Text(-3,-5)[b]{$\to$}
\Text(-3,-12)[b]{$q$}
\Text(93,28)[b]{$\to$}
\Text(93,20)[b]{$q_1$}
\Text(93,-12)[b]{$\to$}
\Text(93,-20)[b]{$q_2$}
%\Text(93,-5)[b]{$\to$}
%\Text(93,-12)[b]{$q$}
%\Text(93,20)[b]{$\to$}
%\Text(93,12)[b]{$p$}
%\Text(52,-30)[b]{$\to$}
%\Text(52,-38)[b]{$p$}
%\Text(-3,-5)[b]{$p$}
\end{axopicture}
}}
\hspace{3mm}
}
\vspace{15mm}
\caption{Two-loop two-point diagram $\hat{I}(q,m_1,...,m_5)$ and three-point diagram $\hat{P}(q,m_1,...,m_6)$ with $q_1^2=q_2^2=0$.
  %which cannot be expressed as a combination of loops and chanins.
%  The examples of usual and dual FIs.
}
\label{sunsetMMm}
\end{figure}
%\vskip 0.5cm

%Note that with the help of this group of transformations (see the appendix) we can not only change the line indices, but also sometimes fold the sum of the product of momenta into one single product.

Below
%in the present analysis
we will concentrate mostly on two-loop two-point and three-point
diagrams,
%particular cases,
which can be taken from 
%Often it’s convenient to calculate complex diagrams.
%use functional relations similar to those obtained in [2, 5], which reduces the amount of computation.
%So, for example, for
the diagram shown in Fig. 1.
We will call them as:
\bea
&&\hat{I}_j = \hat{I}(q,m_j=m \neq 0, m_p=0, p \neq j),~~ \hat{I}_{ij} = \hat{I}(q,m_i=m_j=m \neq 0, m_p=0, p \neq i \neq j), \nonumber \\
&&\hat{I}_{ijs} = \hat{I}(q,m_i=m_j=m_s=m \neq 0, m_p=0, p\neq i \neq j \neq s), \nonumber \\
&&\hat{I}_{ijst} = \hat{I}(q,m_i=m_j=m_s=m_t=m \neq 0, M_p=0, p\neq i \neq j \neq s\neq t), \label{Iijst} \\
&&\hat{P}_j = \hat{P}(q,m_j=m \neq 0, m_p=0, p\neq j),~~ \hat{P}_{ij} = \hat{P}(q,m_i=m_j=m \neq 0, m_p=0, p\neq i \neq j), \nonumber \\
&&\hat{P}_{ijs} = \hat{P}(q,m_i=m_j=m_s=m \neq 0, m_p=0, p\neq i \neq j \neq s), \nonumber \\
&&\hat{P}_{ijst} = \hat{P}(q,m_i=m_j=m_s=m_t=m \neq 0, m_p=0, p\neq i \neq j \neq s\neq t). \label{Iijst}
\eea

Now we repeat once again the procedure of the DE method.
%of differential equations.
%{\bf 2.} Now we will return to the momentum space.
Application of the IBP procedure \cite{Chetyrkin:1981qh}
to loop internal momenta leads to relations between  various FIs
%Feynman integrals
and,  therefore, to the necessity of
calculating only some of them, which in a sense are independent.
These independent diagrams (which were chosen completely arbitrarily, of course) are called master integrals
\cite{Broadhurst:1987ei}.

Applying the IBP procedure \cite{Chetyrkin:1981qh} to the 
master-integrals themselves leads to DEs
%differential equations 
\cite{Kotikov:1990kg,Kotikov:1990zs} for them with
the inhomogeneous terms containing less complex diagrams.
%\footnote{The ``less complex diagrams''  usually contain less number of
%propagators and sometimes they can be represented as diagrams with less
%number of loops and with some ``effective masses'' (see, for example, 
%\cite{FleKoVe,Kniehl:2005bc,Kniehl:2005yc,Fleischer:1999hp} and references therein).}
Applying the IBP procedure to diagrams in inhomogeneous terms leads to new DEs
%differential equations 
for them with new inhomogeneous terms 
containing even more less complex diagrams ($\equiv$ less$^2$ complex ones).
By repeating the procedure several times, in the last step we can obtain 
inhomogeneous terms containing mainly
tadpoles, which can be easily calculated  in-turn.
%(see also the discussions in Section 7 below).

By solving the corresponding DEs
%differential equations
in this last step, the diagrams for the inhomogeneous terms of the DEs
%differential equations
in the previous step can be reproduced.
Repeating the procedure several times, me can get the results for the original Feynman diagram.

Thus, the DE
%differential equation
method procedure is well defined, but it requires a lot of
%handwork
manual work
and a lot of time. So,
the calculations \cite{Fleischer:1997bw} of each of the diagrams $P_6$ and $P_{126}$ took about a month of work
(of course, along with checking the results).
It would be nice, however, to transfer some of the work to the computer. The first attempt based on the properties of the
inverse mass expansion coefficients of the master integrals. It is presented in the next Section. A more modern and efficient
technique is discussed in Section 5.

\section{Evaluation of series}

Calculations of the two-point diagrams shown in Fig. 1, which do not contain elliptic structures, see Fig. 2 in Ref. \cite{FleKoVe},
\footnote{In fact, the results for these two-point diagrams were found in the late eighties and early nineties,
and were planned to be published in a long paper summarizing the results done in Refs. \cite{Kotikov:1990kg,Kotikov:1991hm}.
However, this paper has not been published. These results,
%obtained,
after verification, were published in Ref. \cite{FleKoVe}.}
  as well as calculations of some three-point diagrams shown in Fig. 1, see also Fig. 3 in Ref. \cite{FleKoVe},
lead to results with interesting properties  of their inverse mass expansion coefficients.

%here
%\be
%S_i=S_i(n-1),~~\overline{S}_i=S_i(2n-1)\, . 
%\label{Si(n-1)}
%\ee

\subsection{Properties of series}

%This scheme was successfully used to calculate the two-loop two-point \cite{Kotikov:1990zs,Kotikov:1990kg,Fleischer:1999hp}
%and three-point diagrams \cite{Fleischer:1997bw,FleKoVe,Kniehl:2005bc}
%with one nonzero mass. This procedure is very powerful, but rather complicated. However, there are some simplifications based
%on representations of series of Feynman integrals.

%\begin{figure}[t]
%%\includegraphics[width=0.8\textwidth,height=0.5\textheight,angle=0]{2loop.eps}
%\includegraphics[width=0.6\textwidth,height=0.3\textheight,angle=0]{2loop.eps}
%%\vskip -1.5cm
%%\vspace{-1.5cm}
%%\begin{minipage}[b]{14pc}
%%\caption{
%%%\label{fig:geo2}
%%Two-point diagrams
%%}
%%\end{minipage}
%\end{figure}

%Indeed, the
The inverse-mass expansion of two-loop two-point
%(see Fig. 2)
and three-point diagrams
%(see Fig. 3)
\footnote{The diagrams
  %shown in Figs. 2 and 3,
  are complicated two-loop FIs
  %Feynman integrals
  that do not have cuts of
  three massive particles.  Thus, their results should be expressed as combinations of polylogarithms. 
Note that we consider only three-point diagrams with independent upward momenta $q_1$ and $q_2$, which satisfy 
the conditions $q_1^2=q_2^2=0$ and $(q_1+q_2)^2\equiv q^2 \neq 0$, where $q$ is a downward momentum.}
with one nonzero mass (massless and massive propagators are shown by thinner and thicker 
%dashed and
solid lines,
respectively), can be considered as
\begin{eqnarray}
&&\mbox{ FI} ~ 
%&& 
  = ~
  %\eta K_{\rm FI}
  \, \frac{\hat{N}}{q^{2\alpha}} \,
%\sim ~~
\sum_{n=1} \, C_n \, {(\eta x)}^n
%\frac{{(\eta x)}^n}{n^c}
\, \biggl\{F_0(n) +
\biggl[\ln x
  %  \ln (-x)
  \, F_{1,1}(n)  +
\frac{1}{\varepsilon} \, F_{1,2}(n) \biggr] 
%\nonumber 
\label{FI1} \\
&& + \biggl[\ln^2 x
  %  \ln^2 (-x)
  \, F_{2,1}(n)  + \frac{1}{\varepsilon} \,\ln x
  %\ln (-x)
  \, F_{2,2}(n) + \frac{1}{\varepsilon^2} \, F_{2,3}(n) + \zeta(2)
 \, F_{2,4}(n) \biggr]
%\nonumber \\
%&& + \biggl[ \ln^3 (-x) \, F_{3,1}(n)  + \frac{1}{\varepsilon} \,\ln^2 (-x) \,
% F_{3,2}(n)  + \frac{1}{\varepsilon^2} \,\ln (-x) \,
% F_{3,3}(n)
%+ \frac{1}{\varepsilon^3} \, F_{3,4}(n) \nonumber \\
%&&+ \zeta(2)  \,\ln (-x) \, F_{3,5}(n)
%+ \zeta(3)  \, F_{3,6}(n) \biggr]
%%\nonumber \\
%%&&
+ \cdots \biggr\},
%\label{FI1}
\nonumber
\end{eqnarray}
where $x=q^2/m^2$,
%(as it was shown in (\ref{P126}),
$\eta =1$
%and
or $-1$
%$c =0$, $1$ and $2$,
and $\alpha=1$ and $2$ for
two-point and three-point cases, respectively. 
The normalization factor $\hat{N}={(\overline{\mu}^2/m^2)}^{2\ep}$, where the mass scale $\overline{\mu}=4\pi e^{-\gamma_E} \mu$ is  the standard one of the
$\overline{MS}$-scheme
and $\gamma_E$ is Euler constant.
%is defined below Eq. (\ref{P126}).
%$\hat{N}={(\overline{\mu}^2/m^{2})}^{2\varepsilon}$, 
%where
%where $\overline{\mu}=4\pi e^{-\gamma_E} \mu$ is in the standard
%$\overline{MS}$-scheme and $\gamma_E$ is the Euler constant.
Moreover,
%the space-time dimension is $D=4-2\varepsilon$ and
%\begin{eqnarray}
%C_n ~=~ 1
%%\nonumber
%\label{FI1a}
%\end{eqnarray}
%%with some $c$,
%for diagrams with one-massive-particle-cuts ($m$-cuts) and
%%Here
\begin{eqnarray}
%C_n  ~=~1, ~~~\mbox{ and }~~~
C_n  ~=~ \frac{(n!)^2}{(2n)!} ~\equiv ~ \hat{C}_n
%\nonumber
\label{FI1b}
\end{eqnarray}
%with some $c$,
for diagrams with
%one-massive-particle-cut and
two-massive-particle-cuts ($2m$-cuts). For the diagrams with one-massive-particle-cuts ($m$-cuts)
one has $C_n = 1$.
%, respectively.

For the $m$-cut
%one-massive-particle-cut
case,
the coefficients $F_{N,k}(n)$ should have the form
\begin{eqnarray}
  F_{N,k}(n) ~ \sim ~ \frac{S_{\pm a,...}}{n^b}\, ,
  %~~  \frac{\overline{S}_{\pm a,...}}{n^b}\, ,
  ~~ \frac{\zeta(\pm a)}{n^b}\, , 
%,~~~~(a+b=M-N),
%\nonumber
\label{FI1c}
\end{eqnarray}
where $S_{\pm a,...} \equiv S_{\pm a,...}(j-1)$
%and  $\overline{S}_{\pm a,...} \equiv S_{\pm a,...}(2j-1)$
%$S_{\pm a} \equiv S_{\pm a}(j-1),\ S_{\pm a,\pm b} \equiv
%S_{\pm a,\pm b}(j-1),\ S_{\pm a,\pm b,\pm c} \equiv
%S_{\pm a,\pm b,\pm c}(j-1)$ 
are
% Hereafter we use the following
nested sums \cite{Vermaseren:1998uu}:
\footnote{In our previous papers \cite{Kazakov:1986mu,Kazakov:1987jk,Fleischer:1997bw,FleKoVe}
  the nested sums $K_{a,b,...}(j)=\sum^j_{m=1} \, \frac{(-1)^{m+1}}{m^a} S_{b,...}(m)= - S_{- a,b,...}(j)$ have been used together with their
  analytic continuations \cite{Kazakov:1987jk,Kotikov:1994mi}.}
\be
S_{\pm a}(j)=\sum^j_{m=1} \, \frac{(-1)^m}{m^a},~~ S_{\pm a,\pm b,...}(j)=\sum^j_{m=1} \, \frac{(-1)^m}{m^a} S_{\pm b,...}(m), 
\label{Sin}
\ee
and $\zeta(\pm a) = S_{\pm a}(\infty)$ and $\zeta(\pm a,,\pm b,...) = S_{\pm a,\pm b,...}(\infty)$ are
%Riemann zeta-function.
%harmonic sums defined in (\ref{Sin})
%in (\ref{FI2}).
%\begin{eqnarray}
%\be
%%&&\hspace*{-1cm}
%S_{\pm a}(j)\ =\ \sum^j_{m=1} \frac{(\pm 1)^m}{m^a},
%\ \ S_{\pm a,\pm b,\pm c,\cdots}(j)~=~ \sum^j_{m=1}
%\frac{(\pm 1)^m}{m^a}\, S_{\pm b,\pm c,\cdots}(m),  \label{FI2}
%\ee
%\end{eqnarray}
%and $\zeta(\pm a)$ are
the Euler-Zagier constants.
%\be
%%\begin{eqnarray}
%%&&\hspace*{-1cm}
%\zeta(\pm a)\ =\ \sum^{\infty}_{m=1} \frac{(\pm 1)^m}{m^a},
%\ \ \zeta(\pm a,\pm b,\pm c,\cdots )~=~ \sum^{\infty}_{m=1}
%\frac{(\pm 1)^m}{m^a}\, S_{\pm b,\pm c,\cdots}(m-1),  \label{Euler}
%%\end{eqnarray}
%\ee

%\begin{figure}[t]
%%\includegraphics[width=0.8\textwidth,height=0.60\textheight,angle=0]{9808242v2.threepoint.ps}
%\includegraphics[width=0.6\textwidth,height=0.35\textheight,angle=0]{9808242v2.threepoint.ps}
%%\vskip -1.5cm
%%\vspace{-1.5cm}
%%\begin{minipage}[b]{14pc}
%%\caption{
%%%\label{fig:geo2}
%%Two-point diagrams
%%}
%%\end{minipage}
%\end{figure}

For  $2m$-cut
%two-massive-particle-cut
case,
the coefficients $F_{N,k}(n)$ can be more complicated
%%should have the form
%\footnote{Really, there are even more complicated terms as ones in Eqs.
%(58) and (59) of \cite{FleKoVe}, 
%which come from other $\eta $ values in (\ref{FI1}).
%However,
%%but
%they are outside of our present consideration.}
\begin{eqnarray}
F_{N,k}(n) ~ \sim ~ \frac{S_{\pm a,...}}{n^b},  ~ \frac{V_{a,...}}{n^b}
,  ~ \frac{W_{a,...}}{n^b} \, ,
%\nonumber
\label{FI1d}
\end{eqnarray}
where
%  where
  %$V_{\pm a,...} \equiv V_{\pm a,...}(j-1)$ and 
$W_{\pm a,...} \equiv W_{\pm a,...}(j-1)$
%, $\overline{W}_{\pm a,...} \equiv W_{\pm a,...}(2j-1)$
 and
$V_{\pm a,...} \equiv V_{\pm a,...}(j-1)$
with \cite{FleKoVe}
\begin{eqnarray}
%V_{a}(j)\ =\ \sum^j_{m=1}
%%\, \frac{(m!)^2}{(2m)!}
%\, \frac{\hat{C}_m}{m^a},
%\ \ V_{a,b,c,\cdots}(j)~=~ \sum^j_{m=1}  \,
%%\frac{(m!)^2}{(2m)!} \
%\frac{\hat{C}_m}{m^a}\, S_{b,c,\cdots}(m),  \label{FI4} \\
&&W_{a}(j) ~=~ \sum^j_{m=1} \,
%\frac{(2m)!}{(m!)^2} \,
\frac{\hat{C}_m^{-1}}{m^a},~~~
W_{a,b,c,\cdots}(j)~=~ \sum^j_{m=1} \,
%\frac{(2m)!}{(m!)^2} \
\frac{\hat{C}_m^{-1}}{m^a}\, S_{b,c,\cdots}(m),~~
\label{FI5}\\
%\end{eqnarray}
%$W_{\pm a,...}$ is defined in Eq. (\ref{FI5}) and
%$V_{\pm a,...} \equiv V_{\pm a,...}(j-1)$
%and $W_{\pm a,...} \equiv W_{\pm a,...}(j-1)$
%with
%\begin{eqnarray}
&&V_{a}(j) ~=~ \sum^j_{m=1}
%\, \frac{(m!)^2}{(2m)!}
\, \frac{\hat{C}_m}{m^a},~~~
V_{a,b,c,\cdots}(j)~=~ \sum^j_{m=1}  \,
%\frac{(m!)^2}{(2m)!} \
\frac{\hat{C}_m}{m^a}\, S_{b,c,\cdots}(m),  \label{FI4}
%\\
%W_{a}(j)\ =\ \sum^j_{m=1} \,
%%\frac{(2m)!}{(m!)^2} \,
%\frac{\hat{C}_m^{-1}}{m^a},
%\ \ W_{a,b,c,\cdots}(j)~=~ \sum^j_{m=1}  \,
%%\frac{(2m)!}{(m!)^2} \
%\frac{\hat{C}_m^{-1}}{m^a}\, S_{b,c,\cdots}(m),  \label{FI5}
\end{eqnarray}

The terms $\sim V_{a,...}$ and $\sim W_{a,...}$
can appear only in the case of the $2m$-cut.
%two-massive-particle-cut
%case.
%together with the coefficients $C_n =1$ and
%$C_n = \hat{C}_n
%%(n!)^2/(2n)!
%$, respectively. The terms  $\sim S_{\pm a,...}$
%can appear in combination with
%%come behind the
%both $C_n$ values.
%
The origin of the appearance of these  terms 
%$\sim V_{a,...}$ and
%$\sim W_{a,...}$ in the $2m$-cut
%two-massive-particle-cut case, 
is the product of series (\ref{FI1})
with the different 
%values of the 
coefficients $C_n =1$ and
$C_n = \hat{C}_n
%(n!)^2/(2n)!
$.

\subsection{Two-point examples}

As an example, consider two-loop two-point diagrams $\hat{I}_5$
%,  $I_5$ 
and $\hat{I}_{12}$
%shown in Fig. 2 and  
%considered 
studied in \cite{FleKoVe}

\vskip 0.5cm
\be
\hat{I}_5 \, =
 \hspace{3mm}
%\frac{2^{n-m} \Gamma(n-m+\alpha)}{(n-m)!\Gamma(\alpha)} \, \hspace{3mm}
\raisebox{1mm}{{
    %\begin{picture}(90,30)(0,4)
    \begin{axopicture}(90,10)(0,4)
%  \SetWidth{2.0}
  \SetWidth{0.5}
%\CArc(5,5)(80,20,160)
%\CArc(45,5)(40,0,180)
%\Arc(45,-7)(40,20,160)
\Arc(45,-7)(40,90,160)
%\SetWidth{0.7}
\Arc(45,-7)(40,20,90)
%\Arc[arrow](45,-7)(40,90,160)
\Arc(45,17)(40,200,270)
\SetWidth{2.0}
\Line(45,-25)(45,35)
%\Line[arrow](40,5)(85,5)
%\Arc(45,17)(40,200,270)
\SetWidth{0.5}
\Arc(45,17)(40,270,340)
%\Arc(45,17)(40,200,340)
\Vertex(45,-23){2}
\Vertex(45,33){2}
\Vertex(5,5){2}
\Vertex(85,5){2}
%\SetWidth{1.0}
%\Vertex(5,5){2}
\Line(5,5)(-5,5)
\Line(85,5)(95,5)
%Line(45,-25)(45,-35)
%Line(85,5)(95,10)
%\Vertex(40,5){2}
%\Vertex(40,15){2}
%\Vertex(40,-15){2}
%\Text(25,33)[b]{$m$}
%\Text(63,-10)[b]{$\scriptstyle n$}
%\Text(63,-10)[b]{$n$}
%\Text(33,3)[t]{$\scriptstyle \frac{M^2}{s(1-s)}$}
%\Text(33,1)[t]{$\scriptstyle M^2/[s(1-s)]$}
%\Text(75,-22)[t]{$\alpha$}
%\Text(75,-22)[t]{$\scriptstyle \alpha$}
\Text(-3,-5)[b]{$\to$}
\Text(-3,-12)[b]{$q$}
%\Text(93,-5)[b]{$\to$}
%\Text(93,-12)[b]{$q$}
%\Text(93,20)[b]{$\to$}
%\Text(93,12)[b]{$p$}
%\Text(52,-30)[b]{$\to$}
%\Text(52,-38)[b]{$p$}
%\Text(-3,-5)[b]{$p$}
\end{axopicture}
}}
\hspace{3mm},~~
\hat{I}_{12} \, =
 \hspace{3mm}
%\frac{2^{n-m} \Gamma(n-m+\alpha)}{(n-m)!\Gamma(\alpha)} \, \hspace{3mm}
\raisebox{1mm}{{
    %\begin{picture}(90,30)(0,4)
    \begin{axopicture}(90,10)(0,4)
 \SetWidth{2.0}
 \Arc(45,-7)(40,90,160)
 \Arc(45,17)(40,200,270)
 %\CArc(5,5)(80,20,160)
%\CArc(45,5)(40,0,180)
%\Arc(45,-7)(40,20,160)
 \SetWidth{0.5}
 \Arc(45,-7)(40,20,90)
%\Arc[arrow](45,-7)(40,90,160)
\Line(45,-25)(45,35)
%\Line[arrow](40,5)(85,5)
\Arc(45,17)(40,270,340)
%\Arc(45,17)(40,200,340)
\Vertex(45,-23){2}
\Vertex(45,33){2}
\Vertex(5,5){2}
\Vertex(85,5){2}
%\SetWidth{1.0}
%\Vertex(5,5){2}
\Line(5,5)(-5,5)
\Line(85,5)(95,5)
%Line(45,-25)(45,-35)
%Line(85,5)(95,10)
%\Vertex(40,5){2}
%\Vertex(40,15){2}
%\Vertex(40,-15){2}
%\Text(37,5)[t]{$n$}
%\Text(37,5)[t]{$\scriptstyle n$}
%\Text(33,3)[t]{$\scriptstyle \frac{M^2}{s(1-s)}$}
%\Text(33,1)[t]{$\scriptstyle M^2/[s(1-s)]$}
%\Text(59,5)[t]{$\alpha$}
%\Text(59,5)[t]{$\scriptstyle \alpha$}
\Text(-3,-5)[b]{$\to$}
\Text(-3,-12)[b]{$q$}
%\Text(93,-5)[b]{$\to$}
%\Text(93,-12)[b]{$q$}
%\Text(93,20)[b]{$\to$}
%\Text(93,12)[b]{$p$}
%\Text(52,-30)[b]{$\to$}
%\Text(52,-38)[b]{$p$}
%\Text(-3,-5)[b]{$p$}
\end{axopicture}
}}
\hspace{3mm}
\label{I1+2}
\ee
\vskip 1cm
\noindent
where $\hat{I}_5$ coincides with $I_1(q^2,m^2)$ considered in Section 2.

Their results are
\begin{eqnarray}
\hat{I}_5 &=&  \frac{\hat{N}}{q^{2}} \,
%\sim ~~
\sum_{n=1} \, \frac{x^n}{n} \, \biggl\{
\ln^2 x
%\ln^2 (-x)
- \frac{2}{n} \ln x
%\ln (-x)
+ 2\zeta(2)
+4S_{-2} +2 \frac{2}{n^2} + \frac{2(-)^n}{n^2} \biggr\} \, ,
%\nonumber \\
\label{FI6a} \\
%I_5 &=& \frac{\hat{N}}{q^{2}} \,
%%\sim ~~
%\sum_{n=1} \, \frac{(-x)^n}{n} \, \biggl\{
%- \ln^2 (-x) + \frac{2}{n} \ln (-x)  -2 \zeta(2)
%-4S_{-2} -\frac{2}{n^2} - 2\frac{(-1)^n}{n^2} \biggr\} \, ,
%%\nonumber \\
%\label{FI6b} \\
\hat{I}_{12} &=& - \frac{\hat{N}}{q^{2}} \,
%\sim ~~
\sum_{n=1} \, \frac{(-x)^n}{n^2} \, \biggl\{\frac{1}{n} + \hat{C}_n
%\frac{(n!)^2}{(2n)!}
\, \biggl(
 -2 \ln x
 % \ln (-x)
 -3 W_1 + \frac{2}{n} \biggr) \biggr\} \, .
\label{FI6c}
%\nonumber
\end{eqnarray}

From (\ref{FI6a}) 
%and (\ref{FI6b}) 
one can see that the corresponding functions
$F_{N,k}(n)$ have the form 
\begin{eqnarray}
F_{N,k}(n) ~ \sim ~ \frac{1}{n^{3-N}},~~~~(N\geq 2),
%\nonumber
\label{FI8}
\end{eqnarray}
if we introduce the following complexity of the sums ($\overline{\Phi}=(S,V,W)$) 
%($\eta = \pm$,
%($\sum_{i=1}^m a_i =a$)
\begin{eqnarray}
\overline{\Phi}_{\pm a} \sim \overline{\Phi}_{\pm a_1, \pm a_2}
\sim \overline{\Phi}_{\pm a_1,\pm a_{2},\cdots,\pm a_m}
\sim \zeta_{a} \sim \frac{1}{n^a},~~~~ (\sum_{i=1}^m a_i =a) \, .
\label{FI9}
\end{eqnarray}
%where $\Phi=(S,V,W)$.

The number $3-N$ determines the level of transcendentality (or complexity, or weight)
of the coefficients $F_{N,k}(n)$. The property greatly reduces  the number
of the possible elements in $F_{N,k}(n)$.
The level of transcendentality decreases if we consider the singular parts of diagrams and/or coefficients in front of
$\zeta$-functions and of logarithm powers.
Thus, finding the parts we can predict, the rest is obtained using the ansatz 
based on the results known already, but containing elements with a higher 
level  of transcendentality.

Other two-loop two-point
%$I$-type
integrals in \cite{FleKoVe} have similar form. They were exactly calculated
by DE
%differential equation
method \cite{Kotikov:1990kg,Kotikov:1990zs}. Their representations in the form of Nielsen polylogarithms \cite{Devoto:1983tc}
can be found also in Ref. \cite{FleKoVe}.

\subsection{Three-point examples}

Now we consider two-loop three-point diagrams, 
%three-point ones
%$P_1$, 
$\hat{P}_5$ and  $\hat{P}_{12}$:
%, $P_6$, $P_{13}$ 
%and $\hat{P}_{126}$:
%shown in Fig. 3 and
%calculated in \cite{FleKoVe}
%considered 

\vskip 0.5cm
\be
\hat{P}_{5} \, =
 \hspace{3mm}
%\frac{2^{n-m} \Gamma(n-m+\alpha)}{(n-m)!\Gamma(\alpha)} \, \hspace{3mm}
\raisebox{1mm}
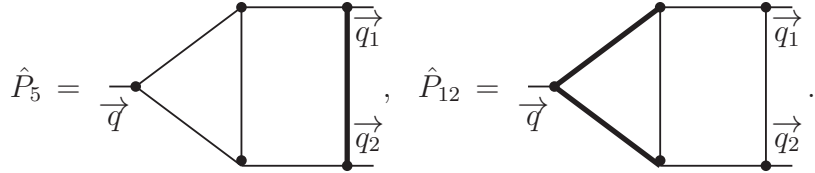
{{
    %\begin{picture}(90,30)(0,4)
    \begin{axopicture}(90,10)(0,4)
%  \SetWidth{2.0}
  \SetWidth{0.7}
%\CArc(5,5)(80,20,160)
%\CArc(45,5)(40,0,180)
%\Arc(45,-7)(40,20,160)
%\Arc(45,-7)(40,20,90)
%%\Arc[arrow](45,-7)(40,90,160)
%\Arc(45,-7)(40,90,160)
  \Line(5,5)(45,35)
  \Line(5,5)(45,-25)
\Line(45,-25)(45,35)
\SetWidth{0.7}
\Line(45,35)(85,35)
\Line(45,-25)(85,-25)
\SetWidth{2.0}
\Line(85,35)(85,-25)
\SetWidth{0.7}
%\Arc(45,17)(40,200,270)
%\Arc[arrow](45,17)(40,270,340)
%\Arc(45,17)(40,200,340)
\Vertex(45,35){2}
\Vertex(45,-23){2}
\Vertex(5,5){2}
\Vertex(85,35){2}
\Vertex(85,-25){2}
%\SetWidth{1.0}
%\Vertex(5,5){2}
\Line(5,5)(-5,5)
\Line(85,35)(95,35)
\Line(85,-25)(95,-25)
%Line(45,-25)(45,-35)
%Line(85,5)(95,10)
%\Vertex(40,5){2}
%\Vertex(40,15){2}
%\Vertex(40,-15){2}
%\Text(25,33)[b]{$m$}
%\Text(63,-10)[b]{$\scriptstyle n$}
%\Text(63,-10)[b]{$n$}
%\Text(33,3)[t]{$\scriptstyle \frac{M^2}{s(1-s)}$}
%\Text(33,1)[t]{$\scriptstyle M^2/[s(1-s)]$}
%\Text(75,-22)[t]{$\alpha$}
%\Text(75,-22)[t]{$\scriptstyle \alpha$}
\Text(-3,-5)[b]{$\to$}
\Text(-3,-12)[b]{$q$}
\Text(93,28)[b]{$\to$}
\Text(93,20)[b]{$q_1$}
\Text(93,-12)[b]{$\to$}
\Text(93,-20)[b]{$q_2$}
%\Text(93,-5)[b]{$\to$}
%\Text(93,-12)[b]{$q$}
%\Text(93,20)[b]{$\to$}
%\Text(93,12)[b]{$p$}
%\Text(52,-30)[b]{$\to$}
%\Text(52,-38)[b]{$p$}
%\Text(-3,-5)[b]{$p$}
\end{axopicture}
}}
\hspace{3mm}
,~~
\hat{P}_{12} \, =
 \hspace{3mm}
%\frac{2^{n-m} \Gamma(n-m+\alpha)}{(n-m)!\Gamma(\alpha)} \, \hspace{3mm}
\raisebox{1mm}{{
    %\begin{picture}(90,30)(0,4)
    \begin{axopicture}(90,10)(0,4)
%  \SetWidth{2.0}
  \SetWidth{2.0}
%\CArc(5,5)(80,20,160)
%\CArc(45,5)(40,0,180)
%\Arc(45,-7)(40,20,160)
%\Arc(45,-7)(40,20,90)
%%\Arc[arrow](45,-7)(40,90,160)
%\Arc(45,-7)(40,90,160)
  \Line(5,5)(45,35)
  \Line(5,5)(45,-25)
\SetWidth{0.7}
  \Line(45,-25)(45,35)
\Line(45,35)(85,35)
\Line(45,-25)(85,-25)
\Line(85,35)(85,-25)
%\Arc(45,17)(40,200,270)
%\Arc[arrow](45,17)(40,270,340)
%\Arc(45,17)(40,200,340)
\Vertex(45,35){2}
\Vertex(45,-23){2}
\Vertex(5,5){2}
\Vertex(85,35){2}
\Vertex(85,-25){2}
%\SetWidth{1.0}
%\Vertex(5,5){2}
\Line(5,5)(-5,5)
\Line(85,35)(95,35)
\Line(85,-25)(95,-25)
%Line(45,-25)(45,-35)
%Line(85,5)(95,10)
%\Vertex(40,5){2}
%\Vertex(40,15){2}
%\Vertex(40,-15){2}
%\Text(25,33)[b]{$m$}
%\Text(63,-10)[b]{$\scriptstyle n$}
%\Text(63,-10)[b]{$n$}
%\Text(33,3)[t]{$\scriptstyle \frac{M^2}{s(1-s)}$}
%\Text(33,1)[t]{$\scriptstyle M^2/[s(1-s)]$}
%\Text(75,-22)[t]{$\alpha$}
%\Text(75,-22)[t]{$\scriptstyle \alpha$}
\Text(-3,-5)[b]{$\to$}
\Text(-3,-12)[b]{$q$}
\Text(93,28)[b]{$\to$}
\Text(93,20)[b]{$q_1$}
\Text(93,-12)[b]{$\to$}
\Text(93,-20)[b]{$q_2$}
%\Text(93,-5)[b]{$\to$}
%\Text(93,-12)[b]{$q$}
%\Text(93,20)[b]{$\to$}
%\Text(93,12)[b]{$p$}
%\Text(52,-30)[b]{$\to$}
%\Text(52,-38)[b]{$p$}
%\Text(-3,-5)[b]{$p$}
\end{axopicture}
}}
\hspace{3mm} \, .
\nonumber
\ee
\vskip 1cm

Their results are (see \cite{FleKoVe}):
%which have been calculated in [FKV98]
\begin{eqnarray}
%P_1 &=& \frac{\hat{N}}{(q^{2})^2} \,
%%\sim ~~
%\sum_{n=1} \, \frac{x^n}{n} \, \biggl\{
%-\frac{1}{2\ep^3} - \frac{S_1}{\ep^2} + \frac{1}{2\ep}
%\biggl[5S_2-S_1^2+ \frac{2}{n^2} - \frac{2}{n} \ln (-x)  
%%\nonumber \\ && 
%+\frac{1}{2} \ln^2 (-x) - \zeta(2) \biggr]
%\nonumber \\ &&
%-\frac{8}{3}\zeta_3 -\biggl(S_1+\frac{1}{n}\biggr)\zeta_2
%+ \frac{8}{3} S_3 + \frac{9}{2}S_1S_2 + \frac{5}{6}S_1^3 + 4 \frac{S_2}{n}+
%2\frac{S_1}{n^2} + \frac{3}{n^3} \nonumber \\
%&&+ \biggl(\zeta_2-4S_2-2\frac{S_1}{n}- \frac{3}{n^2}\biggr)\ln (-x)
%+ \biggl(S_1 + \frac{3}{2n}\biggr)\ln^2 (-x) -  \frac{1}{2}\ln^3 (-x)
%\biggr\} \, ,
%\label{FI7a} \\
\hat{P}_5 &=& \frac{\hat{N}}{(q^{2})^2} \,
%\sim ~~
\sum_{n=1} \, \frac{x^n}{n} \, \biggl\{
-6\zeta_3 + 2S_1\zeta_2
+6S_3-2S_1S_2+ 4 \frac{S_2}{n}-\frac{S_1^2}{n}
+ 2\frac{S_1}{n^2} \nonumber \\
&&+ \biggl(-4S_2+S_1^2-2\frac{S_1}{n}\biggr) \, \ln x
%\ln (-x)
+ S_1 \ln^2 x
%\ln^2 (-x)
\biggl\} \, ,
\label{FI7b} \\
%P_6 &=& \frac{\hat{N}}{(q^{2})^2} \,
%\sum_{n=1} \, \frac{(-x)^n}{n} \, \biggl\{
%- \frac{1}{\ep^2} \biggl[\ln(-x)-\frac{1}{n}\biggr]
%+ \frac{1}{\ep}
%\biggl[\zeta_2-3S_2-4S_{-2} - 3 \frac{S_1}{n}
%- \frac{3}{n^2}  \nonumber \\
%&&+ \biggl(3S_1+ \frac{3}{n}\biggr) \ln (-x)
%-\frac{3}{2}\ln^2 (-x) \biggr]
%%\nonumber \\&&
%+2\zeta_3 + \biggl(7S_1+\frac{2}{n}\biggr)\zeta_2
%-2S_3-9S_1S_2
%\nonumber \\&&
%+10S_{-3}-12S_{-2,1}-4S_1S_{-2} - \frac{7}{2}\frac{S_2}{n}
%-\frac{9}{2}\frac{S_1^2}{n}
%-5\frac{S_1}{n^2} -\frac{7}{n^3}
%%\nonumber \\ &&
%+ \biggl(\frac{7}{2}S_2-\frac{9}{2}S_1^2
%\nonumber \\&&
%+5\frac{S_1}{n}+\frac{7}{n^2}
%-2\zeta_2\biggr)\ln (-x)
%+  \frac{1}{2}\biggl(7S_1+\frac{7}{n}\biggr)\ln^2 (-x)
%+\frac{7}{6}\ln^3 (-x)\biggl\} \, ,
%\label{FI7c} \\
%P_{13} &=& \frac{\hat{N}}{(q^{2})^2} \,
%%\sim ~~
%\sum_{n=1} \, x^n \,
%%\frac{x^n}{n} \,
%\biggl\{
%-\frac{S_2}{2\ep^2} - \frac{1}{2\ep}
%\biggl[S_3+4S_{1,2} -4 \frac{S_2}{n}  \biggr]  +\frac{S_2}{2}\zeta_2
%\nonumber \\
%&& -S_{1,3}-3S_{3,1}+3S_{1,1,2}+3S_{1,2,1}-S_2^2
%+\biggl(7S_3 -8S_{1,2}\biggr) S_1 + \frac{5}{2}S_1^2 S_2
%\biggr\} \, ,
%\label{FI7d} \\
\hat{P}_{12} &=& \frac{\hat{N}}{(q^{2})^2} \,
%\sim ~~
\sum_{n=1} \, \frac{(-x)^n}{n^2} \, \hat{C}_n 
%\frac{(n!)^2}{(2n)!}
\, \biggl\{
 \frac{2}{\ep^2} + \frac{2}{\ep} \biggl(S_1 -3 W_1 + \frac{1}{n} -\ln x
 %\ln (-x)
\biggr) -6 W_2 -18 W_{1,1}
\nonumber \\ && -13S_2 + S_1^2- 6S_1W_1 +2 \frac{S_1}{n} +
\frac{2}{n^2}
-2 \bigg(S_1+\frac{1}{n}\biggr)\ln x
%\ln (-x)+ \ln^2 (-x)
+ \ln^2 x 
\biggr\} \, ,
\label{FI7e}
\end{eqnarray}

 Now the coefficients
$F_{N,k}(n)$ have the form
\begin{eqnarray}
  %\frac{1}{n^{c}} \,
  F_{N,k}(n) ~ \sim ~ \frac{1}{n^{4-N}},~~~~(N\geq 3),
%\nonumber
\label{FI12}
\end{eqnarray}

The diagram 
%$P_1$, 
$P_5$ 
%and $P_6$ 
(and also $P_1$, $P_3$, $P_6$ and $P_{126}$ in \cite{FleKoVe})
was calculated
exactly by differential equation
method \cite{Kotikov:1990kg,Kotikov:1990zs}.
\footnote{The evaluation of the inverse mass expansion coefficients  is demonstrated in Ref. \cite{Kotikov:2020ccc}.}
To find the results for  
%$P_{13}$ and 
$P_{12}$ (and also
%for
all others
in \cite{FleKoVe}) we have used the knowledge of the several $n$ terms in the
inverse-mass expansion (\ref{FI1}) (usually less than $n=100$) and
%plausible arguments:\\
the following arguments: 
%%(see \cite{FleKoVe1} and discussions therein):
\begin{itemize}
\item
%{\bf 2.}~
If
%there is
a two-loop two-point diagram with a ``similar topology'' (for example, 
%$I_1$ for $P_1$ and $P_3$,  $I_5$ for $P_5$ and $P_6$, 
$I_{12}$ for $P_{12}$, etc.)
was already calculated, we should consider a similar set of basic elements for corresponding $F_{N,k}(n)$ of
two-loop three-point diagrams
but with a higher level of complexity.
\item
%{\bf 3.}~
Let the diagram under consideration contain singularities and/or powers of logarithms.
Since the coefficients are very simple before the leading singularity, or the largest degree of the logarithm, or
the largest $\zeta$-function, they can often be predicted directly from the first few terms of the expansion.

Moreover, often we can calculate the singular part using a different technique
(see \cite{FleKoVe} for extraction of $\sim W_1(n)$ part). Then we should expand the
singular parts, find the main elements and try to use them
(with the corresponding increase in the level of complexity) in order to predict
the regular part of the diagram. If we need to find  $\ep$-suppressed terms, we should
increase the level of complexity of the corresponding basic elements.
\end{itemize}

Later, using the ansatz for $F_{N,k}(n)$ and several terms (usually less than 100) in the above
expression, which can be  exactly calculated, we obtain a system of
algebraic equations for the parameters of the ansatz. Solving the system, we
can obtain the analytical results for FIs
%Feynman integrals
without exact calculations.
To check the results, we only need to calculate a few more terms in the
above inverse-mass expansion (\ref{FI1}) and compare them with the
predictions of our anzatz with the fixed coefficients indicated above.

Thus, the considered arguments give a possibility to find results for many complicated
two-loop three-point diagrams without direct calculations.
Several process options have been successfully used to calculate
Feynman diagrams for many processes (see 
\cite{Fleischer:1997bw,FleKoVe,Kotikov:2020ccc,Kniehl:2005bc,Kniehl:2005yc,Kniehl:2006bg}).

Note that properties similar to (\ref{FI8}) and (\ref{FI12})
%were
  %recently
%observed \cite{Eden:2012rr} also in the so-called double
%operator-product-expansion limit of some four-point diagrams.
%These diagrams
%are encoded the quantum corrections to the four-point correlator and
%have been considered in \cite{Eden:2012rr} up-to three-loop level of accuracy.\\
%Moreover, property (\ref{FI9}) at
but $b=0$ in (\ref{FI1c})
was found for the eigenvalues of anomalous dimensions \cite{KL}
and coefficient functions \cite{Bianchi:2013sta}, as well as in the next-to-leading corrections \cite{KL00} to
the BFKL equation \cite{BFKL} for $N=4$ the Super Yang-Mils (SYM) model. Such a strong restriction made it
possible to obtain anomalous dimensions in the first three orders of the perturbation theory directly
from the corresponding results for QCD (the "most complicated" parts are the same in $N=4$ SYM and QCD)
\cite{KoLiVe,KLOV},
as well as in the 4th, 5th, 6th and 7th orders (see \cite{KLRSV},  \cite{LuReVe}, 
\cite{Marboe:2014sya} and \cite{Marboe:2016igj}, respectively) in the algebraic Bethe ansatz \cite{Staudacher:2004tk}.

Note that the series (\ref{FI6a}), (\ref{FI6c}) and (\ref{FI7b}) can be expressed as a combination of the Nilson \cite{Devoto:1983tc}
 and Remiddi-Vermaseren \cite{Remiddi:1999ew}
 polylogarithms with weight $4-N$ (see \cite{FleKoVe,Fleischer:1997bw}). More complicated cases were examined in
 \cite{Davydychev:2003mv}.

\subsection{Properties of massive diagrams}
%\subsection{Modern technique of massive diagrams}

%{\bf 3.}~
%\item
Coefficients of the inverse-mass-series expansions of the two-point and three-point FIs have the structure
(\ref{FI8}) and (\ref{FI12}) with the rule
(\ref{FI9}). Note that these conditions greatly reduce the number of
possible harmonic sums. In turn, the restriction is associated with a DE specific form
%of differential equations
for the considered FIs.
%Feynman integrals under consideration.
The DEs
%Differential equations
can be formally represented
%master integrals,
as \cite{Kotikov:2010gf,Kotikov:2012ac} (see the example $I_1(q^2,m^2)$ considered in section 2)
\bea
\left((x+a)\frac{d}{dx} - \overline{k}(x)\ep
\right) \, \mbox{ FI } \, = \,
 \mbox{ less complicated diagrams} (\equiv \rm{FI}_1),
%\nonumber
\label{Int}
 \eea
with some number $a$ and some function $\overline{k}(x)$.
This form is generated by IBP procedure for diagrams including
an inner $\hat{n}$-leg one-loop subgraph, which in turn contains the product $k^{\mu_1}...k^{\mu_m}$ of its internal momenta $k$ with
$m=n-3$.

Indeed, for ordinary degrees 
  $\alpha_i=1+a_i\ep$ with arbitrary $a_i$ of subgraph propagators, the IBP relation (\ref{TreIBPM}) gives the
  coefficient
%  \footnote{See, for example, Eqs. (1) and (23) in the first and second papers of Ref. \cite{Kazakov:1986mu}.
%    Please note that these results were made in coordinate space and can be applied for dual diagrams (see discussion
%    in Introduction). The
%    results in the momentum space are the same. we also note that for an inner loop corresponding to the case $n=2$,
%    it is convenient to take the index $\alpha_2=2+a_2\ep$.}
  $d-2\alpha_1 - \sum^p_{i=2} \alpha_i+m \sim \ep$ for $m=n-3$. Important examples of applying the
  rule are the diagrams $\hat{I}_5$, $\hat{I}_{12}$ and $\hat{P}_5$, $\hat{P}_{12}$
  %, $P_{126}$
  (for the case $n=2$ and $n=3$)
  %  in Fig. 2 and planar ones in Fig. 3 (for the case $n=3$)
  and also the diagrams 
 % \footnote{
    %    I thanks Rutger Boels for him information about Ref.  \cite{Gehrmann:2011xn}.}
    in Ref.
  \cite{Gehrmann:2011xn} (for the case $n=3$ and $n=4$).
  %\footnote{
  However, we note that the results for the non-planar diagrams (see Fig. 3 of \cite{FleKoVe}) obey the Eq. (\ref{FI12}) but
  their subgraphs do not comply with the above rule. 
  %  are not in agreement with the above rule.
The disagreements may be related to the on-shall vertex of the subgraph, but this requires additional research.
 
Taking the set of less complicated Feynman integrals $\rm{FI}_1$ as diagrams having internal $\hat{n}$-leg subgraphs,
we get their result structure similar to the one given above (\ref{FI12}), but with  a lower level of complexity.

So, the integrals $\rm{FI}_1$ should obey to the following equation (see $J_2^{(1)}(q^2,m^2)$ in Appendix A)
%similar to one in (\ref{Int}).
%It has the following form
\bea
\left((x+a_1)\frac{d}{dx} - \overline{k}_1(x)\ep
\right) \, \mbox{ FI$_1$ } \, = \,
 \mbox{ less$^2$ complicated diagrams} (\equiv \rm{FI}_2) .
%\nonumber
\label{Int.1}
 \eea

 Thus, we will have the a set of equations for all Feynman integrals $\rm{FI}_n$ as
 \bea
\left((x+a_n)\frac{d}{dx} - \overline{k}_n(x)\ep
\right) \, \mbox{ FI$_n$ } \, = \,
 \mbox{ less$^{n+1}$ complicated diagrams} (\equiv \rm{FI}_{n+1}),
%\nonumber
\label{Int.n}
 \eea
 with the last integral $\rm{FI}_{n+1}$ contains only tadpoles.
 Note that for the case $n=2$ the diagrams corresponding for the example $I_1(q^2,m^2)$, satisfy the system of equations,
 formally represented as eq. (\ref{Int.n}).

 %as a tidepole.
  
\section{Modern technique of massive diagrams}

In the last decade, several popular applications of DEs
%differential equations
have emerged, allowing the use of computer resources and
thus to obtain results for very complicated FIs.

In my opinion, the most successfully used approach are the so-called the canonical form representation \cite{Henn:2013pwa}
of DEs
%differential equations
(and its generalizations in Refs. \cite{Adams:2018yfj,Adams:2016xah}) ,
the method \cite{Papadopoulos:2014lla} of simplified DEs,
%differential equations,
and the ability to use the effective mass (see
eq. (\ref{loopM})), as well as their combinations. DEs
%Differential equations
are also effectively used in calculating FIs with an
elliptical structure (see \cite{Weinzierl:2019pfw}).

% Moreover, following  \cite{Henn:2013pwa}, we can reconstruct the above set of inhomogeneous equations as 
%the the homogeneous matrix equation
%\footnote{For complicated diagrams, there is an extension in Ref. \cite{Adams:2018yfj}.}
%(see Ref. \cite{Lee:2014ioa} containg methods to obtain the equation)

\subsection{Canonical form of differential equations}

In our notation (see eqs. (\ref{Int}) - (\ref{Int.n})), the canonical form \cite{Henn:2013pwa}, which was
introduced by Johannes Henn in 2013 and is widely popular now (there is a huge number of publications, which simply cannot be listed
here), represents a homogeneous matrix equation of the form (see also the review \cite{Argeri:2007up})
%\footnote{For complicated diagrams, there is an extension in Ref. \cite{Adams:2018yfj}.}
\begin{eqnarray} 
  \frac{d}{dx} \widehat{FI} - \ep  \widehat{K}(x)  \widehat{FI}=0,
%    \nonumber
\label{Henn}
\end{eqnarray}
 for the vector
 \begin{eqnarray}
   \widehat{FI} =
   \left(  \begin{array}{l}
     \rm{FI}\\  \rm{FI}_1/\ep \\   ... \\  \rm{FI}_n/\ep^n
\end{array} \right)
   \,,
   %~~~ \frac{d}{dx} \widehat{FI} - \ep  \widehat{K}  \widehat{FI}=0,
   \nonumber
\end{eqnarray}
 where the matrix $ \widehat{K}$ contains
 the functions $\overline{k}_j/(x+a_j)$ as its elements.
 The form (\ref{Henn}) is called as the ``canonic basic''.
 %It is now very popular 
 %(see, for example, recent papers in Ref. \cite{Duhr:2020kzd}).
% report \cite{Henn:2014qga} and discussion therein).

 Note that obtaining it is far from trivial (see, for example, Appendix A for $\rm{FI}_{n=2}$ diagrams). Moreover, it is not always
 achievable (see \cite{Adams:2018yfj,Adams:2016xah}),
where FIs were considered that are not reducible to (\ref{Henn})), and to obtain it is sometimes associated with a
nontrivial analysis (see Refs. \cite{Lee:2014ioa} and \cite{Lee:2017oca} containing methods and criterion to obtain the equation, respectively).
However, the form of (\ref{Henn}) is very
convenient as it can be easily diagonalized. Note that formally 
%Please note that
for real calculations of $\rm{FI}_{n}$ it is convenient to replace
 \begin{eqnarray}
   %   \rm{FI}_{n} \to \widetilde{\rm{FI}}_{n},~~~
   \rm{FI}_{n} =  \widetilde{\rm{FI}}_{n}  \overline{\rm{FI}}_{n},
%   \left(x+a_n\right)^{\overline{k}_n \ep},
   \nonumber
\end{eqnarray}
 where the term $\overline{\rm{FI}}_{n}$ obeys the corresponding homogeneous equation
 \bea
\left((x+a_n)\frac{d}{dx} - \overline{k}_n(x)\ep
\right) \, \overline{\rm{FI}}_{n}
%\mbox{ FI$_n$ }
\, =\, 0,
% \mbox{ less$^{n+1}$ complicated diagrams} (\equiv \rm{FI}_{n+1}),
%\nonumber
\label{IntBar.n}
\eea
 
The replacement simplifies the above equation (\ref{Int.n}) to the following form
\bea
(x+a_n)\frac{d}{dx}  \,  \widetilde{\rm{FI}}_{n} \, = \,   \widetilde{\rm{FI}}_{n+1}
\frac{ \overline{\rm{FI}}_{n+1}}{ \overline{\rm{FI}}_{n}} \, ,
%\frac{ \left(x+a_{n+1}\right)^{\overline{k}_{n+1} \ep}}{ \left(x+a_n\right)^{\overline{k}_n \ep}},
% \mbox{ less$^{n+1}$ complicated diagrams} (\equiv \rm{FI}_{n+1}),
%\nonumber
\label{Int.n.si}
 \eea
 having the solution
 %in the form of Goncharov Polylogariths
 \bea
 \widetilde{\rm{FI}}_{n}(x) = \int^x_0 \frac{dx_1}{x_1+a_{n}} \widetilde{\rm{FI}}_{n+1}(x_1)
%\frac{ \left(y+a_{n+1}\right)^{\overline{k}_{n+1} \ep}}{ \left(k+a_n\right)^{\overline{k}_n \ep}}
\frac{ \overline{\rm{FI}}_{n+1}(x_1)}{ \overline{\rm{FI}}_{n}(x_1)} \, .
 \label{Int.n.si1}
 \eea

 %{\bf 3.}
Usually there are some cancellations in the ratio
$\overline{\rm{FI}}_{n+1}/ \overline{\rm{FI}}_{n}$
and sometimes it is equal to 1. In the last case, the
equation (\ref{Int.n.si1}) coincides with the definition of Goncharov Polylogariths \cite{Goncharov:2001iea}
(see also the review \cite{Duhr:2014woa} and
 the references therein).

 Sometimes the integrand in (\ref{Int.n.si1}) can have a quadratic form in the denominator, for example, $x_1^2\pm x_1+1$
 (sign $\pm$ can change, including when passing from the Euclidean metric to the Minkowski metric).
Such forms appeared in two-point FIs, $\hat{I}_{14}$, $\hat{I}_{15}$ and $\hat{I}_{123}$ and can be represented as
Nilson three-logarithm with complicated argument, i.e. $\rm{Li}_3(-y^3)$, where $y=(\sqrt{x+4}-x)/(\sqrt{x+4}+x)$ is so-called
conformal variable,
as well as in the transform in \cite{Kotikov:2007vr} of $H(-r,...$) functions, introduced in \cite{Aglietti:2003yc}, to
the Remiddi-Vermaseren polylopagitms \cite{Remiddi:1999ew}
of variable $\sim y$  where one integral representation contains the factor $x_1^2\pm x_1+1$ in the denominator and is thus left in
this form.
Terms of this kind have appeared recently in \cite{Lee:2021iid} also and could be shown
to be mapped into cyclotomic harmonic polylogarithms \cite{Ablinger:2011te} in Ref. \cite{ABS}.
We note  that such terms come also in contributions of the massive form factors at 3-loop order \cite{Lee:2016ixa}.
 Already before, the study of such integral representations
leads to the discovery of cyclotomic Polylogarithms, see \cite{Ablinger:2011te} and Ref. \cite{Blumlein:2018cms} for a review. \\

\subsection{Other approaches}

Here we will consider other methods that can be connected both with each other and with the canonical form and its generalizations.
Unfortunately, we cannot pretend here to be complete in listing all the approaches.\footnote{A short review of many approaches has
recently been presented as an introduction to this volume \cite{Blu.2103}.}

{\bf 1.}~~The {\it simplified DE
  %differential equations
  approach} \cite{Papadopoulos:2014lla} is based on violation of momentum
conservation by the parameter $x$,
with some propagator. Using the IBP relations, we can obtain set of equations which depend on $x$. We can solve it with the boundary conditions at $x=0$
and take the limit $x \to 1$. The equations in this approach are usually representable in canonical form, which leads to very important results (see
\cite{Canko:2020ylt}). 

{\bf 2.}~~ {\it Series expansions in singular and regular fixed points} \cite{Lee:2017qql} (see also Ref. \cite{Lee:2020zfb} and discussion therein)
for DE systems,
%of differential equations,
which generate eq. (\ref{Henn}), for example, as
\be
\ep  \widehat{K}(x) \to \widehat{K}_1(x) + \ep  \widehat{K}_2(x) \, .
\label{gHenn}
\ee
The results are obtained in the form of
%generalized
Goncharov polylogarithms \cite{Goncharov:2001iea} and, in some complicated cases, numerically.

{\bf 3.}~~ {\it Symmetries of FIs}
%Feynman Integrals}
  is a general method introduced in \cite{Kol:2015gsa} which
  associates with any given Feynman diagram a system of partial DEs.
  %differential equations.
  The method uses the same variations which are used in the DE
  method
  %of Differential Equations
  \cite{Kotikov:1990kg} and IBP technique
  %Integration By Parts one
  \cite{Chetyrkin:1981qh}, but distinguishes itself by associating
with any diagram a natural Lie group which acts on the diagram's parameter space. This approach was further developed and
numerous diagrams have been analyzed within it (see the recent paper \cite{Kol:2020tqt} and discussions and references therein).

{\bf 4.}~~ Using {\it the effective mass} (\ref{loopM}) reduces the number of loops in the considered diagram. In the cases
under consideration, two-loop
diagrams were reduced to one-loop ones. Then, one-loop diagrams were easily calculated using the DE method,
%of differential equations,
and the required two-loop diagrams were presented as integrals of the obtained one-loop results (see Ref. \cite{Kniehl:2005bc}). 

\subsection{Elliptic structure}

%It is a well-known fact that while many Feynman integrals admit  representations  in  terms  of  so-called  multiple  Polylogarithms  (MPLs),
%this  space  offunctions  is  not  sufficient  to  express  integrals  when  the  number  of  physical  scales  is  sufficiently large.
%More
Recently, the scientific community has centered its attention to the study of FIs
%Feynman integrals
whose geometric properties are defined
by elliptic curves.
%So, we
We already have a lot of progress in understanding simplest functions beyond  usual polylogarithms,  the  so-called  elliptic
polylogarithms
%(for a review,
(see the recent papers \cite{Weinzierl:2019pfw,Vanhove:2018mto,Broedel:2018rwm,Walden:2020odh} and references and discussions therein).
%Unfortunately, this subject is beyond of the present consideration but we would like to note only about some integral representations, which
%can be used together with  elliptic Polylogarithms.
Unfortunately, this topic is beyond the scope of this consideration (discussions about elliptic polylogarithms can be found in Ref.
\cite{Weinzierl:2019pfw}, which is a contribution to this Volume), but we would like to point out only some of the integral representations that
can be used in conjunction with elliptic polylogarithms or even instead of elliptic polylogarithms.

%Using
{\it The effective mass}
%This
form (\ref{loopM}) turned out to be convenient for integrals containing an elliptic structure, since it allows one to
represent the final result (see Ref. \cite{Kniehl:2005bc})
as an integral containing an elliptic kernel (i.e., a root of a polynomial of the 3rd or 4th degree) and a remainder
represented in the form of an ordinary (Goncharov) polylogarithms. This approach can be an alternative to the introduction of
elliptic polylogarithms, which have a very complex structure (see, for example, the recent paper \cite{Campert:2020yur},
where the study of sunsets in special kinematics was carried out both in the form of elliptic polylogarithms
(following Ref. \cite{Broedel:2017kkb}), as well as in the form of integral representations containing an
elliptic kernel and ordinary polylogarithms. Notice, that 
%where
such analysis has been
done in  all orders of the dimensional regulator following the corresponding results in Ref.  \cite{Kalmykov:2008ge}).

At the end of
%To finish
the section, we would like to note about the
%In a
recent paper \cite{Bezuglov:2020ywm}, where the results for the most complex two-point single-mass diagrams containing an elliptical structure
were obtained in the following form: using the effective mass representation, the original FIs were presented as integrals
of  one-loop diagrams dependent on the ratio $\mu/m$. These one-loop diagrams were considered in a generalized canonical form (\ref{gHenn}).
%(i.e. $k(x)\ep \to k_1(x)+ k_2(x)\ep$ in eq. (\ref{Henn})).
The authors of Ref.  \cite{Bezuglov:2020ywm}
have obtained very convenient representations
for extremely complicated FIs.

\section{Conclusion}

In this short review we examined the applicability of DEs
%of differential equations
for calculating FIs.
We have considered an example $I_1(q^2,m^2)$, which led to the DE method sometime ago.
%that led at one time to the DE method.
%of differential equations.
The consistent application of IBP relations to $I_1(q^2,m^2)$, and then to the diagrams of the inhomogeneous terms
that arise each time, made it possible to obtain a DE hierarchy
%of differential equations
for increasingly
simple diagrams obtained at each step by reducing one propagator.
As noted in section 3.1, the DE method is well defined but requires a lot of manual work and a lot of time. 

Next, we showed an effective method restoring the exact result for two-point and three-point two-loop diagrams
in terms of inverse-mass-expansion coefficients, which have a beautiful structure and can be predicted using
the corresponding coefficients at the poles or at transcendental constants 
%transcending constants
such as Euler's $\zeta$-functions.
These predictions were verified by analytical calculations of the first few terms using computer programs.
Thus, this method is, apparently, the first, where computer programs were used for FI calculations using differential equations.

We have also given a brief overview of modern popular techniques such as the `{\it canonical form} of DEs
%differential equations}
\cite{Henn:2013pwa}, the {\it simplified DE
  %differential equations
  approach} \cite{Papadopoulos:2014lla} and the
method of the {\it effective mass}, see, for example, Ref.  \cite{Kniehl:2005yc}. Section 5.2 lists other popular approaches
as well. 

The canonical form \cite{Henn:2013pwa}, and its generalizations \cite{Adams:2018yfj,Adams:2016xah}, are probably the most
commonly used approaches (at least as a part of the calculations.  

The effective mass method (see \cite{Kniehl:2005yc}) allows one to actually work with diagrams that have fewer loops than
the original ones.
The results for the original diagrams are obtained in the form of integral representations, where the integrand expressions are
determined by calculating the diagrams with fewer loops. So, in Ref. \cite{Kniehl:2005bc} the two-loop diagrams with an elliptic
structure were considered. The corresponding one-loop diagrams depending on the effective mass {\it have no elliptical
  structure}. Thus, the results of the original diagrams were presented in the form of integral representations containing an
elliptic kernel (i.e., a root of a polynomial of the 3rd or 4th degree) and ordinary polylogarithms. These representations can
be used instead of elliptic polylogarithms, and even more complex objects than elliptic polylogarithms, see
\cite{Bezuglov:2020ywm} and discussions therein.

Following the discussion in Section 5.3, the combined application of the {\it effective-mass} approach and
generalizations of the canonical form for effective-mass-dependent diagrams can yield results for very complicated FIs.
Such an analysis has already been carried out in the recent article \cite{Bezuglov:2020ywm} and, in our opinion, similar
calculations can be performed in the near future for many complicated FIs.\\
 
The author 
%A.V.K. 
thanks Johannes Blumlein for invitation to present a contribution to the Proceedings of International Conference  
"Antidifferentiation and the Calculation of Feynman Amplitudes" (4-9 October 2020, Zeuthen, Germany) and Andrey Pikelner
for help with {\bf Axodraw2} \cite{Collins:2016aya}. The author thanks Kay Schonwald also for a strong improvement in article style.

%           APPENDICES

%\appendix
\section{Appendix. Massive part of $J_1(q^2,m^2)$ in eq. (\ref{eq4}).}
%\section{Exact results for one-loop integrals}
\label{sec:a}
\def\theequation{A\arabic{equation}}
\setcounter{equation}0

In this appendix we consider the following diagrams
\vskip 0.5cm
\be
I_2^{(\alpha)}(q^2,m^2) \, = \,
\hspace{3mm}
  \raisebox{1mm}{{
    %\begin{picture}(90,30)(0,4)
    \begin{axopicture}(90,10)(0,4)
%  \SetWidth{2.0}
  \SetWidth{0.5}
%\CArc(5,5)(80,20,160)
%\CArc(45,5)(40,0,180)
\Arc(45,-7)(40,20,160)
  %\Line[arrow](5,5)(40,5)
%\Line[arrow](40,5)(85,5)
\SetWidth{1.5}
\Arc(5,-35)(40,20,85)
\SetWidth{0.5}
\Arc(45,17)(40,200,270)
\SetWidth{0.5}
\Arc(45,17)(40,270,340)
\SetWidth{0.5}
\Vertex(5,5){2}
\Vertex(85,5){2}
\Vertex(43,-23){2}
%\SetWidth{1.0}
%\Vertex(5,5){2}
\Line(5,5)(-5,5)
\Line(85,5)(95,5)
%\Vertex(40,5){2}
%\Vertex(40,15){2}
%\Vertex(40,-15){2}
%\Text(65,-10)[b]{$n$}
%\Text(65,10)[b]{$m$}
%\Text(33,3)[t]{$\scriptstyle \frac{M^2}{s(1-s)}$}
%\Text(33,1)[t]{$\scriptstyle M^2/[s(1-s)]$}
%\Text(65,-25)[t]{$\alpha-1$}
\Text(33,7)[t]{$\alpha$}
%\Text(25,-20)[t]{$2$}
\Text(-3,-5)[b]{$\to$}
\Text(-3,-12)[b]{$q$}
%\Text(-3,-5)[b]{$p$}
\end{axopicture}
  }}
  \hspace{3mm} \, ,~~
S^{(\beta,\alpha)}(q^2,m^2) \, = \,
  \hspace{3mm}
  \raisebox{1mm}{{
    %\begin{picture}(90,30)(0,4)
    \begin{axopicture}(90,10)(0,4)
%  \SetWidth{2.0}
  \SetWidth{0.5}
%\CArc(5,5)(80,20,160)
%\CArc(45,5)(40,0,180)
\Arc(45,-7)(40,20,160)
\Arc(45,17)(40,200,270)
\Arc(45,17)(40,270,340)
 \SetWidth{1.5}
\Line(5,5)(82,5)
 \SetWidth{0.5}
%\Line[arrow](40,5)(85,5)
\SetWidth{1.5}
%\Arc(5,-35)(40,20,85)
\SetWidth{0.5}
%\Arc(45,17)(40,200,270)
%\Arc(45,17)(40,300,270)
\SetWidth{0.5}
%\Arc(45,17)(40,270,340)
\SetWidth{0.5}
\Vertex(5,5){2}
\Vertex(85,5){2}
%\Vertex(43,-23){2}
%\SetWidth{1.0}
%\Vertex(5,5){2}
\Line(5,5)(-5,5)
\Line(85,5)(95,5)
%\Vertex(40,5){2}
%\Vertex(40,15){2}
%\Vertex(40,-15){2}
%\Text(65,-10)[b]{$n$}
%\Text(65,10)[b]{$m$}
%\Text(33,3)[t]{$\scriptstyle \frac{M^2}{s(1-s)}$}
%\Text(33,1)[t]{$\scriptstyle M^2/[s(1-s)]$}
%\Text(65,-25)[t]{$\alpha-1$}
\Text(40,18)[t]{$\alpha$}
\Text(40,-8)[t]{$\beta$}
\Text(-3,-5)[b]{$\to$}
\Text(-3,-12)[b]{$q$}
%\Text(-3,-5)[b]{$p$}
\end{axopicture}
  }}
  \hspace{3mm} \, .
  \label{eqJ1}
\ee
\vskip 1cm

The IBP relations for the internal loop of the diagram produce two equations:
%\vskip 0.5cm
\bea
&&(d-1-2\alpha) \, I_2^{(\alpha)}(q^2,m^2)  = \alpha \, J_2^{(\alpha+1)}(q^2,m^2) - m^2 \, \alpha \, I_2^{(\alpha+1)}(q^2,m^2)  \, ,
  \label{eq1J1}\\
%  &&\nonumber \\ && \nonumber \\
  %&& \nonumber \\&& \nonumber \\
%&& \nonumber \\
  &&(d-3) \, I_2^{(1)}(q^2,m^2)  =
  T_{0,2}(m^2=0) \, L_{1,1}(q^2) \, - S^{(2,1)}(q^2,m^2)
%\hspace{3mm}
  \nonumber \\ && \nonumber \\
  %&& \nonumber \\
  %&& \nonumber \\
&& \nonumber \\
&&\hspace{2.5cm} 
  - m^2 \, \hspace{3mm}
  \raisebox{1mm}{{
    %\begin{picture}(90,30)(0,4)
    \begin{axopicture}(90,10)(0,4)
%  \SetWidth{2.0}
  \SetWidth{0.5}
%\CArc(5,5)(80,20,160)
%\CArc(45,5)(40,0,180)
\Arc(45,-7)(40,20,160)
  %\Line[arrow](5,5)(40,5)
%\Line[arrow](40,5)(85,5)
\SetWidth{1.5}
\Arc(5,-35)(40,20,85)
\SetWidth{0.5}
\Arc(45,17)(40,200,270)
\SetWidth{0.5}
\Arc(45,17)(40,270,340)
\SetWidth{0.5}
\Vertex(5,5){2}
\Vertex(85,5){2}
\Vertex(43,-23){2}
%\SetWidth{1.0}
%\Vertex(5,5){2}
\Line(5,5)(-5,5)
\Line(85,5)(95,5)
%\Vertex(40,5){2}
%\Vertex(40,15){2}
%\Vertex(40,-15){2}
%\Text(65,-10)[b]{$n$}
%\Text(65,10)[b]{$m$}
%\Text(33,3)[t]{$\scriptstyle \frac{M^2}{s(1-s)}$}
%\Text(33,1)[t]{$\scriptstyle M^2/[s(1-s)]$}
%\Text(65,-25)[t]{$\alpha-1$}
%\Text(33,7)[t]{$2$}
\Text(25,-20)[t]{$2$}
\Text(-3,-5)[b]{$\to$}
\Text(-3,-12)[b]{$q$}
%\Text(-3,-5)[b]{$p$}
\end{axopicture}
  }}
  \hspace{3mm}
- 2m^2 \, I_2^{(2)}(q^2,m^2) \, , 
  \label{eq1J2}
  \eea
  \vskip 1cm
  where
%\vskip 1cm
\be
 J_2^{(\alpha)}(q^2,m^2) =
  T_{0,\alpha}(m^2) \, L_{1,1}(q^2) \, - S^{(1,2)}(q^2,m^2)
\, .
   \label{J2a}
  \ee
%  \vskip 1cm
%  where

We note that
%Moreover,
$T_{0,2}(m^2=0)=0$ in dimensional regularization and
  \be
T_{0,2}(m^2) L_{1,1}(q^2) =  \frac{1}{(4\pi)^{d}} \, \frac{R(0,2)A(1,1)}{m^{2(2-d/2)}q^{2(2-d/2)}} \, ,
\label{T02L11}
\ee
where $R(\alpha_1,\alpha_2)$ and $A(\alpha_1,\alpha_2)$ are given in eqs. (\ref{R}) and (\ref{A}), respectively.

The IBP relations for internal triangles of the diagram $I_2^{(1)}(q^2,m^2)$ produce two additional equations:
%\vskip 1cm
\bea
&&(d-4) \, I_2^{(1)}(q^2,m^2) = S^{(2,1)}(q^2,m^2) -  J_2^{(2)}(q^2,m^2) - m^2 \,  I_2^{(2)}(q^2,m^2)
  %- T_2(m^2) \, L_{1,1}(q^2) 
  %\hspace{3mm}
\nonumber \\ && \nonumber \\ && \nonumber \\
%&& \nonumber \\
%&& \nonumber \\
&& \hspace{3.5cm} 
%- T_2(m^2) \, L_{1,1}(q^2) 
%- m^2 \,  I_2^{(2)}(q^2,m^2) 
-q^2 \,  \hspace{3mm}
  \raisebox{1mm}{{
    %\begin{picture}(90,30)(0,4)
    \begin{axopicture}(90,10)(0,4)
%  \SetWidth{2.0}
  \SetWidth{0.5}
%\CArc(5,5)(80,20,160)
%\CArc(45,5)(40,0,180)
\Arc(45,-7)(40,20,160)
  %\Line[arrow](5,5)(40,5)
%\Line[arrow](40,5)(85,5)
\SetWidth{1.5}
\Arc(5,-35)(40,20,85)
\SetWidth{0.5}
\Arc(45,17)(40,200,270)
\SetWidth{0.5}
\Arc(45,17)(40,270,340)
\SetWidth{0.5}
\Vertex(5,5){2}
\Vertex(85,5){2}
\Vertex(43,-23){2}
%\SetWidth{1.0}
%\Vertex(5,5){2}
\Line(5,5)(-5,5)
\Line(85,5)(95,5)
%\Vertex(40,5){2}
%\Vertex(40,15){2}
%\Vertex(40,-15){2}
%\Text(65,-10)[b]{$n$}
%\Text(65,10)[b]{$m$}
%\Text(33,3)[t]{$\scriptstyle \frac{M^2}{s(1-s)}$}
%\Text(33,1)[t]{$\scriptstyle M^2/[s(1-s)]$}
%\Text(65,-25)[t]{$\alpha-1$}
\Text(45,30)[t]{$2$}
%\Text(25,-20)[t]{$2$}
\Text(-3,-5)[b]{$\to$}
\Text(-3,-12)[b]{$q$}
%\Text(-3,-5)[b]{$p$}
\end{axopicture}
  }}
  \hspace{3mm}
  \, ,
  \label{eq1J3}\\
  &&\nonumber \\ && \nonumber \\ && \nonumber \\
  %&& \nonumber \\
%&& \nonumber \\
&&(d-4) \, \hspace{3mm}
  \raisebox{1mm}{{
    %\begin{picture}(90,30)(0,4)
    \begin{axopicture}(90,10)(0,4)
%  \SetWidth{2.0}
  \SetWidth{0.5}
%\CArc(5,5)(80,20,160)
%\CArc(45,5)(40,0,180)
\Arc(45,-7)(40,20,160)
  %\Line[arrow](5,5)(40,5)
%\Line[arrow](40,5)(85,5)
\SetWidth{0.5}
\Arc(5,-35)(40,20,85)
\SetWidth{1.5}
\Arc(45,17)(40,200,270)
\SetWidth{0.5}
\Arc(45,17)(40,270,340)
\SetWidth{0.5}
\Vertex(5,5){2}
\Vertex(85,5){2}
\Vertex(43,-23){2}
%\SetWidth{1.0}
%\Vertex(5,5){2}
\Line(5,5)(-5,5)
\Line(85,5)(95,5)
%\Vertex(40,5){2}
%\Vertex(40,15){2}
%\Vertex(40,-15){2}
%\Text(65,-10)[b]{$n$}
%\Text(65,10)[b]{$m$}
%\Text(33,3)[t]{$\scriptstyle \frac{M^2}{s(1-s)}$}
%\Text(33,1)[t]{$\scriptstyle M^2/[s(1-s)]$}
%\Text(65,-25)[t]{$\alpha-1$}
%\Text(33,7)[t]{$2$}
%\Text(25,-20)[t]{$2$}
\Text(-3,-5)[b]{$\to$}
\Text(-3,-12)[b]{$q$}
%\Text(-3,-5)[b]{$p$}
\end{axopicture}
  }}
  \hspace{3mm} = S^{(2,1)}(q^2,m^2) - T_2(m^2=0) \, L_{1,1}(q^2) 
  %\hspace{3mm}
\nonumber \\ && \nonumber \\ && \nonumber \\&& \nonumber \\
%&& \nonumber \\
&&\hspace{1.5cm} 
%- T_2(m^2) \, L_{1,1}(q^2) 
+ m^2 \, \hspace{3mm}
  \raisebox{1mm}{{
    %\begin{picture}(90,30)(0,4)
    \begin{axopicture}(90,10)(0,4)
%  \SetWidth{2.0}
  \SetWidth{0.5}
%\CArc(5,5)(80,20,160)
%\CArc(45,5)(40,0,180)
\Arc(45,-7)(40,20,160)
  %\Line[arrow](5,5)(40,5)
%\Line[arrow](40,5)(85,5)
\SetWidth{0.5}
\Arc(5,-35)(40,20,85)
\SetWidth{1.5}
\Arc(45,17)(40,200,270)
\SetWidth{0.5}
\Arc(45,17)(40,270,340)
\SetWidth{0.5}
\Vertex(5,5){2}
\Vertex(85,5){2}
\Vertex(43,-23){2}
%\SetWidth{1.0}
%\Vertex(5,5){2}
\Line(5,5)(-5,5)
\Line(85,5)(95,5)
%\Vertex(40,5){2}
%\Vertex(40,15){2}
%\Vertex(40,-15){2}
%\Text(65,-10)[b]{$n$}
%\Text(65,10)[b]{$m$}
%\Text(33,3)[t]{$\scriptstyle \frac{M^2}{s(1-s)}$}
%\Text(33,1)[t]{$\scriptstyle M^2/[s(1-s)]$}
%\Text(65,-25)[t]{$\alpha-1$}
\Text(33,7)[t]{$2$}
%\Text(25,-20)[t]{$2$}
\Text(-3,-5)[b]{$\to$}
\Text(-3,-12)[b]{$q$}
%\Text(-3,-5)[b]{$p$}
\end{axopicture}
  }}
  \hspace{3mm}
-q^2 \,  \hspace{3mm}
  \raisebox{1mm}{{
    %\begin{picture}(90,30)(0,4)
    \begin{axopicture}(90,10)(0,4)
%  \SetWidth{2.0}
  \SetWidth{0.5}
%\CArc(5,5)(80,20,160)
%\CArc(45,5)(40,0,180)
\Arc(45,-7)(40,20,160)
  %\Line[arrow](5,5)(40,5)
%\Line[arrow](40,5)(85,5)
\SetWidth{1.5}
\Arc(5,-35)(40,20,85)
\SetWidth{0.5}
\Arc(45,17)(40,200,270)
\SetWidth{0.5}
\Arc(45,17)(40,270,340)
\SetWidth{0.5}
\Vertex(5,5){2}
\Vertex(85,5){2}
\Vertex(43,-23){2}
%\SetWidth{1.0}
%\Vertex(5,5){2}
\Line(5,5)(-5,5)
\Line(85,5)(95,5)
%\Vertex(40,5){2}
%\Vertex(40,15){2}
%\Vertex(40,-15){2}
%\Text(65,-10)[b]{$n$}
%\Text(65,10)[b]{$m$}
%\Text(33,3)[t]{$\scriptstyle \frac{M^2}{s(1-s)}$}
%\Text(33,1)[t]{$\scriptstyle M^2/[s(1-s)]$}
%\Text(65,-25)[t]{$\alpha-1$}
\Text(45,30)[t]{$2$}
%\Text(25,-20)[t]{$2$}
\Text(-3,-5)[b]{$\to$}
\Text(-3,-12)[b]{$q$}
%\Text(-3,-5)[b]{$p$}
\end{axopicture}
  }}
  \hspace{3mm}
  \, .
%  \, ,
  \label{eq1J4}
\eea
\vskip 1cm
%with $T_2(m^2=0)=0$.

Using Eqs. (\ref{eq1J2}) and (\ref{eq1J4}) as the combination: $2 \times$(\ref{eq1J2}) + (\ref{eq1J4}) , we have
\vskip 0.5cm
\bea &&
%\be
(3d-10) \, I_2^{(1)}(q^2,m^2) = -4m^2 \, I_2^{(2)}(q^2,m^2)
%  \nonumber \\ && \nonumber \\ && \nonumber \\&& \nonumber \\
%&&
%\hspace{2.5cm} 
 -m^2 \, \hspace{3mm}
  \raisebox{1mm}{{
    %\begin{picture}(90,30)(0,4)
    \begin{axopicture}(90,10)(0,4)
%  \SetWidth{2.0}
  \SetWidth{0.5}
%\CArc(5,5)(80,20,160)
%\CArc(45,5)(40,0,180)
\Arc(45,-7)(40,20,160)
  %\Line[arrow](5,5)(40,5)
%\Line[arrow](40,5)(85,5)
\SetWidth{0.5}
\Arc(5,-35)(40,20,85)
\SetWidth{1.5}
\Arc(45,17)(40,200,270)
\SetWidth{0.5}
\Arc(45,17)(40,270,340)
\SetWidth{0.5}
\Vertex(5,5){2}
\Vertex(85,5){2}
\Vertex(43,-23){2}
%\SetWidth{1.0}
%\Vertex(5,5){2}
\Line(5,5)(-5,5)
\Line(85,5)(95,5)
%\Vertex(40,5){2}
%\Vertex(40,15){2}
%\Vertex(40,-15){2}
%\Text(65,-10)[b]{$n$}
%\Text(65,10)[b]{$m$}
%\Text(33,3)[t]{$\scriptstyle \frac{M^2}{s(1-s)}$}
%\Text(33,1)[t]{$\scriptstyle M^2/[s(1-s)]$}
%\Text(65,-25)[t]{$\alpha-1$}
\Text(33,7)[t]{$2$}
%\Text(25,-20)[t]{$2$}
\Text(-3,-5)[b]{$\to$}
\Text(-3,-12)[b]{$q$}
%\Text(-3,-5)[b]{$p$}
\end{axopicture}
  }}
  \hspace{3mm}
  \nonumber \\ && \nonumber \\ && \nonumber \\
  %&& \nonumber \\
&& \hspace{3.5cm} 
  -q^2 \,  \hspace{3mm}
  \raisebox{1mm}{{
    %\begin{picture}(90,30)(0,4)
    \begin{axopicture}(90,10)(0,4)
%  \SetWidth{2.0}
  \SetWidth{0.5}
%\CArc(5,5)(80,20,160)
%\CArc(45,5)(40,0,180)
\Arc(45,-7)(40,20,160)
  %\Line[arrow](5,5)(40,5)
%\Line[arrow](40,5)(85,5)
\SetWidth{1.5}
\Arc(5,-35)(40,20,85)
\SetWidth{0.5}
\Arc(45,17)(40,200,270)
\SetWidth{0.5}
\Arc(45,17)(40,270,340)
\SetWidth{0.5}
\Vertex(5,5){2}
\Vertex(85,5){2}
\Vertex(43,-23){2}
%\SetWidth{1.0}
%\Vertex(5,5){2}
\Line(5,5)(-5,5)
\Line(85,5)(95,5)
%\Vertex(40,5){2}
%\Vertex(40,15){2}
%\Vertex(40,-15){2}
%\Text(65,-10)[b]{$n$}
%\Text(65,10)[b]{$m$}
%\Text(33,3)[t]{$\scriptstyle \frac{M^2}{s(1-s)}$}
%\Text(33,1)[t]{$\scriptstyle M^2/[s(1-s)]$}
%\Text(65,-25)[t]{$\alpha-1$}
\Text(45,30)[t]{$2$}
%\Text(25,-20)[t]{$2$}
\Text(-3,-5)[b]{$\to$}
\Text(-3,-12)[b]{$q$}
%\Text(-3,-5)[b]{$p$}
\end{axopicture}
  }}
  \hspace{3mm}
  \, .
%  \, ,
  \label{eq1J5}
\eea
\vskip 1cm

So, we have for the mass-dependent part of $J_1(q^2,m^2)$, see Eq. (\ref{eq4}),
\vskip 0.5cm
%\bea&&
\be
m^2 \, \hspace{3mm}
  \raisebox{1mm}{{
    %\begin{picture}(90,30)(0,4)
    \begin{axopicture}(90,10)(0,4)
%  \SetWidth{2.0}
  \SetWidth{0.5}
%\CArc(5,5)(80,20,160)
%\CArc(45,5)(40,0,180)
\Arc(45,-7)(40,20,160)
  %\Line[arrow](5,5)(40,5)
%\Line[arrow](40,5)(85,5)
\SetWidth{0.5}
\Arc(5,-35)(40,20,85)
\SetWidth{1.5}
\Arc(45,17)(40,200,270)
\SetWidth{0.5}
\Arc(45,17)(40,270,340)
\SetWidth{0.5}
\Vertex(5,5){2}
\Vertex(85,5){2}
\Vertex(43,-23){2}
%\SetWidth{1.0}
%\Vertex(5,5){2}
\Line(5,5)(-5,5)
\Line(85,5)(95,5)
%\Vertex(40,5){2}
%\Vertex(40,15){2}
%\Vertex(40,-15){2}
%\Text(65,-10)[b]{$n$}
%\Text(65,10)[b]{$m$}
%\Text(33,3)[t]{$\scriptstyle \frac{M^2}{s(1-s)}$}
%\Text(33,1)[t]{$\scriptstyle M^2/[s(1-s)]$}
%\Text(65,-25)[t]{$\alpha-1$}
\Text(33,7)[t]{$2$}
%\Text(25,-20)[t]{$2$}
\Text(-3,-5)[b]{$\to$}
\Text(-3,-12)[b]{$q$}
%\Text(-3,-5)[b]{$p$}
\end{axopicture}
  }}
  \hspace{3mm}
+q^2 \,  \hspace{3mm}
  \raisebox{1mm}{{
    %\begin{picture}(90,30)(0,4)
    \begin{axopicture}(90,10)(0,4)
%  \SetWidth{2.0}
  \SetWidth{0.5}
%\CArc(5,5)(80,20,160)
%\CArc(45,5)(40,0,180)
\Arc(45,-7)(40,20,160)
  %\Line[arrow](5,5)(40,5)
%\Line[arrow](40,5)(85,5)
\SetWidth{1.5}
\Arc(5,-35)(40,20,85)
\SetWidth{0.5}
\Arc(45,17)(40,200,270)
\SetWidth{0.5}
\Arc(45,17)(40,270,340)
\SetWidth{0.5}
\Vertex(5,5){2}
\Vertex(85,5){2}
\Vertex(43,-23){2}
%\SetWidth{1.0}
%\Vertex(5,5){2}
\Line(5,5)(-5,5)
\Line(85,5)(95,5)
%\Vertex(40,5){2}
%\Vertex(40,15){2}
%\Vertex(40,-15){2}
%\Text(65,-10)[b]{$n$}
%\Text(65,10)[b]{$m$}
%\Text(33,3)[t]{$\scriptstyle \frac{M^2}{s(1-s)}$}
%\Text(33,1)[t]{$\scriptstyle M^2/[s(1-s)]$}
%\Text(65,-25)[t]{$\alpha-1$}
\Text(45,30)[t]{$2$}
%\Text(25,-20)[t]{$2$}
\Text(-3,-5)[b]{$\to$}
\Text(-3,-12)[b]{$q$}
%\Text(-3,-5)[b]{$p$}
\end{axopicture}
  }}
  \hspace{3mm}
%  \nonumber \\ && \nonumber \\ && \nonumber \\&& \nonumber \\
%&&
%\hspace{2.5cm} 
  =
  \left[3d-10 - 4m^2 \frac{d}{dm^2}\right] \,
I_2^{(1)}(q^2,m^2) \, ,
    \label{eq1J6}
\ee
\vskip 0.5cm
\noindent
i.e. the mass-dependent combinations is expressed through the diagram $I_2^{(1)}(q^2,m^2)$ and its derivative.

Using eq. (\ref{eq1J1}), one obtains
%with $\alpha=1$, we have
%\vskip 1cm
\be
\left[d-2-\alpha - m^2 \frac{d}{dm^2}\right] \, I_2^{(\alpha)}(q^2,m^2) = \alpha \, J_2^{(\alpha+1)}(q^2,m^2) 
 \, ,
     \label{eq1J7}
\ee
%\vskip 1cm
i.e. the diagram $I_2^{(\alpha)}(q^2,m^2)$ obeys the differential equation with the inhomogeneous term $J_2^{(\alpha+1)}(q^2,m^2)$ having very simple form: it contains
only one-loop diagrams.
%To look it, w
We see that the last term in $J_2^{(\alpha)}(q^2,m^2)$, see Eq. (\ref{J2a}), is expressed through  massive one loop
$M_{\alpha_1,\alpha_2}(q^2,m^2)$:
\vskip 0.5cm
\be
  M_{\alpha_1,\alpha_2}(q^2,m^2)= \int \frac{Dk}{(q-k)^{2\alpha_1}{(k^2+m^2)}^{\alpha_2}} 
= \hspace{3mm}
\raisebox{1mm}{{
    %\begin{picture}(90,30)(0,4)
    \begin{axopicture}(90,10)(0,4)
%  \SetWidth{2.0}
  \SetWidth{0.5}
%\CArc(5,5)(80,20,160)
%\CArc(45,5)(40,0,180)
\Arc(45,-7)(40,20,160)
  %\Line[arrow](5,5)(40,5)
%\Line[arrow](40,5)(85,5)
\SetWidth{1.5}
\Arc(45,17)(40,200,340)
\SetWidth{0.5}
\Vertex(5,5){2}
\Vertex(85,5){2}
%\SetWidth{1.0}
%\Vertex(5,5){2}
\Line(5,5)(-5,5)
\Line(85,5)(95,5)
%\Vertex(40,5){2}
%\Vertex(40,15){2}
%\Vertex(40,-15){2}
\Text(45,-16)[b]{$\alpha_2$}
%\Text(65,10)[b]{$m$}
%\Text(33,3)[t]{$\scriptstyle \frac{M^2}{s(1-s)}$}
%\Text(33,1)[t]{$\scriptstyle M^2/[s(1-s)]$}
\Text(45,27)[t]{$\alpha_1$}
%\Text(45,-29)[t]{$\alpha_2$}
\Text(-3,-5)[b]{$\to$}
\Text(-3,-12)[b]{$q$}
%\Text(-3,-5)[b]{$p$}
\end{axopicture}
}}
\hspace{3mm} \, .
%= \frac{1}{(4\pi)^{d/2}} \, \frac{A(\alpha_1,\alpha_2)}{q^{2(\alpha_1+\alpha_2-d/2)}} \, ,
\label{Mp}
\ee
\vskip 1cm

Indeed,
%The last diagram
%is really one-loop massive diagram
%\vskip 1cm
\be
S^{(1,\alpha)}(q^2,m^2)  =
= \, A(1,1)
%\frac{A(1,1)}{(4\pi)^{d/2}} \,
\, M_{2-d/2,\alpha}(q^2,m^2)
%\frac{A(\alpha_1,\alpha_2)}{q^{2(\alpha_1+\alpha_2-d/2)}}
\, .
\label{Last}
\ee
%\vskip 0.5cm
%\ee
The one-loop diagram $M_{2-d/2,\alpha}(q^2,m^2)$
%which
%where the final one-loop massive diagrams
can be evaluated by one of some effective methods, for example, by Feynman parameters.

We would like to note that $I_2^{(1)}(q^2,m^2)$ satisfies eq. (\ref{eq1J7}) with $\alpha=1$
that is not of the type of (\ref{Int}). But the integral  $I_2^{(2)}(q^2,m^2)$  satisfies eq. (\ref{eq1J7}) with $\alpha=2$
%that
and is of the type of (\ref{Int}). So, it is convenient to rewrite (\ref{eq1J6}) with $I_2^{(2)}(q^2,m^2)$ in its r.h.s.:
\vskip 0.5cm
%\bea&&
\be
m^2 \, \hspace{3mm}
  \raisebox{1mm}{{
    %\begin{picture}(90,30)(0,4)
    \begin{axopicture}(90,10)(0,4)
%  \SetWidth{2.0}
  \SetWidth{0.5}
%\CArc(5,5)(80,20,160)
%\CArc(45,5)(40,0,180)
\Arc(45,-7)(40,20,160)
  %\Line[arrow](5,5)(40,5)
%\Line[arrow](40,5)(85,5)
\SetWidth{0.5}
\Arc(5,-35)(40,20,85)
\SetWidth{1.5}
\Arc(45,17)(40,200,270)
\SetWidth{0.5}
\Arc(45,17)(40,270,340)
\SetWidth{0.5}
\Vertex(5,5){2}
\Vertex(85,5){2}
\Vertex(43,-23){2}
%\SetWidth{1.0}
%\Vertex(5,5){2}
\Line(5,5)(-5,5)
\Line(85,5)(95,5)
%\Vertex(40,5){2}
%\Vertex(40,15){2}
%\Vertex(40,-15){2}
%\Text(65,-10)[b]{$n$}
%\Text(65,10)[b]{$m$}
%\Text(33,3)[t]{$\scriptstyle \frac{M^2}{s(1-s)}$}
%\Text(33,1)[t]{$\scriptstyle M^2/[s(1-s)]$}
%\Text(65,-25)[t]{$\alpha-1$}
\Text(33,7)[t]{$2$}
%\Text(25,-20)[t]{$2$}
\Text(-3,-5)[b]{$\to$}
\Text(-3,-12)[b]{$q$}
%\Text(-3,-5)[b]{$p$}
\end{axopicture}
  }}
  \hspace{3mm}
+q^2 \,  \hspace{3mm}
  \raisebox{1mm}{{
    %\begin{picture}(90,30)(0,4)
    \begin{axopicture}(90,10)(0,4)
%  \SetWidth{2.0}
  \SetWidth{0.5}
%\CArc(5,5)(80,20,160)
%\CArc(45,5)(40,0,180)
\Arc(45,-7)(40,20,160)
  %\Line[arrow](5,5)(40,5)
%\Line[arrow](40,5)(85,5)
\SetWidth{1.5}
\Arc(5,-35)(40,20,85)
\SetWidth{0.5}
\Arc(45,17)(40,200,270)
\SetWidth{0.5}
\Arc(45,17)(40,270,340)
\SetWidth{0.5}
\Vertex(5,5){2}
\Vertex(85,5){2}
\Vertex(43,-23){2}
%\SetWidth{1.0}
%\Vertex(5,5){2}
\Line(5,5)(-5,5)
\Line(85,5)(95,5)
%\Vertex(40,5){2}
%\Vertex(40,15){2}
%\Vertex(40,-15){2}
%\Text(65,-10)[b]{$n$}
%\Text(65,10)[b]{$m$}
%\Text(33,3)[t]{$\scriptstyle \frac{M^2}{s(1-s)}$}
%\Text(33,1)[t]{$\scriptstyle M^2/[s(1-s)]$}
%\Text(65,-25)[t]{$\alpha-1$}
\Text(45,30)[t]{$2$}
%\Text(25,-20)[t]{$2$}
\Text(-3,-5)[b]{$\to$}
\Text(-3,-12)[b]{$q$}
%\Text(-3,-5)[b]{$p$}
\end{axopicture}
  }}
  \hspace{3mm}
%  \nonumber \\ && \nonumber \\ && \nonumber \\&& \nonumber \\
%&&
%\hspace{2.5cm} 
  =
  \frac{3d-10}{d-3} \, J_2^{(2)}(q^2,m^2) - \frac{d-2}{d-3} \,
  m^2 \, I_2^{(2)}(q^2,m^2) \, .
    \label{eq1J6a}
\ee
\vskip 0.5cm
%i.e. the mass-dependent combinations is expressed through the diagram $I_2^{(1)}(q^2,m^2)$ and its derivative.

Now we should compare the IBP-based equations for $J_2^{(2)}(q^2,m^2)$ and $J_2^{(3)}(q^2,m^2)$ obtained in the right-hand sides of (\ref{eq1J6a}) and
(\ref{eq1J7}), respectively, with Eq. (\ref{Int}).
Since $J_2^{(3)}(q^2,m^2)=-(d/dm^2) \, J_2^{(2)}(q^2,m^2)$, consider only $J_2^{(2)}(q^2,m^2)$. 

So, we should prepare   the IBP-based equations for the massive one-loop diagrams $M_{\ep,2}(q^2,m^2)$ and $M_{\ep,3}(q^2,m^2)$.
%, which contribute to r.h.s.
%of eqs. (\ref{eq1J6a}) and   (\ref{eq1J7}) with $\alpha=2$ and compare them with eq. (\ref{Int}).
%
Applying IBP procedure with massive distinguished line to $M_{\ep,2}(q^2,m^2)$, we have
\be
(-3\ep) M_{\ep,2}(q^2,m^2) = \ep \bigl[M_{1+\ep,1}(q^2,m^2) - (q^2+m^2) \, M_{1+\ep,2}(q^2,m^2)\bigr] -4m^2 \, M_{\ep,3}(q^2,m^2) \, .
  \label{Mep2}
\ee

The corresponding applications of the IBP procedure with massless distinguished line to $M_{1+\ep,1}(q^2,m^2)$ and $M_{1+\ep,2}(q^2,m^2)$
lead to the following results:
\bea
&&(1-4\ep) M_{1+\ep,1}(q^2,m^2) = M_{\ep,2}(q^2,m^2) - (q^2+m^2) \, M_{1+\ep,2}(q^2,m^2) \, ,
\label{M1ep1}\\
&&-4\ep M_{1+\ep,2}(q^2,m^2) = 2 \, M_{\ep,3}(q^2,m^2) - 2 \, (q^2+m^2) \, M_{1+\ep,3}(q^2,m^2) \, ,
\label{M1ep2}
\eea

%Since $2M_{1+\ep,3}(q^2,m^2)=(-d/(dm^2)) \, M_{1+\ep,2}(q^2,m^2)$, the
The last equations has the following form
\be
\left[-4\ep -  (q^2+m^2) \,\frac{d}{dm^2} \, \right] \, M_{1+\ep,2}(q^2,m^2) = - \,\frac{d}{dm^2} \, M_{\ep,2}(q^2,m^2)
\label{M1ep2.1}
\ee

Putting (\ref{M1ep1}) to (\ref{Mep2}), we have after little algebra
\be
-4\ep(1-3\ep) M_{\ep,2}(q^2,m^2) = -2\ep(1-4\ep)\, (q^2+m^2) \, M_{1+\ep,2}(q^2,m^2)\bigr] -4(1-4\ep)\,m^2 \, M_{\ep,3}(q^2,m^2) \, ,
 \label{Mep2.1}
\ee
which transforms to
\be
\left[-4\ep(1-3\ep) - 2(1-4\ep)\, \frac{d}{dm^2} \, \right] \, M_{\ep,2}(q^2,m^2) = -2\ep(1-4\ep)\, (q^2+m^2) \, M_{1+\ep,2}(q^2,m^2)
\label{Mep2.2}
\ee

So, eqs. (\ref{M1ep2.1}) and (\ref{Mep2.2}) can be frustrated as a system of equations having a form similar to equation (\ref{Int}).

\end{document}